\documentclass[aps,pre,reprint,showpacs,floatfix]{revtex4-1}
\usepackage{graphicx}
\usepackage[applemac]{inputenc}
\newcommand\abs[1]{\left|{#1}\right|}
\newcommand\Inline[3]{\begin{figure}[th]\begin{center}\includegraphics[width=3in]{#1}
\caption{\label{#2}#3}\end{center}\end{figure}}
\newcommand\Figure[3]{\begin{figure}[th]\begin{center}
\caption{\label{#2}#3}\end{center}\end{figure}}
\renewcommand\Figure[3]{}
\def\,{,\!}
\def\top{_{\rm top}}
\begin{document}
\title{A dynamic model of time-dependent complex networks}
\author{Scott A. Hill}
\affiliation{Department of Physics, University of Toledo, Toledo OH 43606}
\author{Dan Braha}
\affiliation{New England Complex Systems Institute, Cambridge MA 02138, and University of Massachusetts}
\date{\today}
\begin{abstract}
The characterization of the ``most connected'' nodes in static or slowly evolving complex networks has helped in understanding and predicting the behavior of social, biological, and technological networked systems, including their robustness against failures, vulnerability to deliberate attacks, and diffusion properties.  However, recent empirical research of large dynamic networks (characterized by irregular connections that evolve rapidly) has demonstrated that there is little continuity in degree centrality of nodes over time, even when their degree distributions follow a power law.  This unexpected dynamic centrality suggests that the connections in these systems are not driven by preferential attachment or other known mechanisms.
We present a novel approach to explain real-world dynamic networks and qualitatively reproduce these dynamic centrality phenomena. This approach is based on a dynamic preferential attachment mechanism, which exhibits a sharp transition from a base pure random walk scheme.
\end{abstract}

\maketitle
\newcommand\eg{\emph{e.g., }}

\section{Introduction}
The modern study of social networks largely presupposes static or slowly evolving complex networks, as clearly indicated by an extensive review of the pertinent literature \cite{WF99, CSW7, AB2, ASB0, BLM6, CRT7}.  In the social and behavioral sciences, this static perspective of networks has generated a wide spectrum of formal representations and tools for measuring structural and locational properties of social networks (\eg centrality and prestige), social network roles and positions (\eg structural equivalence), and local (subgraph) structures.  Statistical models of static networks are prevalent \cite{WF99, CSW7}.  The study of networks has been developed beyond the social sciences by a substantial number of papers that have extended the study of universal properties in physical systems to complex networks in social, biological, and technological systems; these recent advances have generated an extensive research effort in understanding the effect of static connectivity patterns on various dynamical processes occurring on top of networks \cite{BBV8}.  One of the most important concepts in complex network studies is \emph{network centrality}, which characterizes the ``most connected'' nodes in static complex networks~\cite{WF99, CSW7}.  Among other things, the concept of network centrality has led to the study of ``scale-free networks''~\cite{BA99}, where the \emph{degree centrality} of nodes is distributed according to a power-law or other long-tail distribution, implying the existence of highly-connected nodes called \emph{hubs} \cite{ASB0,BA99}.  The leading explanation for this uneven distribution of connectivity, present in many real networks, is based on growth and preferential attachment mechanisms, in which the structural changes of real networks are governed by the dynamical evolution of the system (see~\cite{BA99} and its many variants~\cite{DM2}). The long-tail characterization of complex networks has helped in understanding and predicting the behavior of social, biological, and technological networked systems, including their robustness against failures~\cite{AJB0,CEB0}, vulnerability to deliberate attacks~\cite{AJB0,CEB1}, and diffusion properties~\cite{BBV8,PV2a}. This static network centrality perspective has also led the discussion of strategies for immunizing particular individuals to reduce disease propagation~\cite{DB2,PV2b,CHB3,GEM6}, advertising to ``opinion leaders''~\cite{R3}, and targeting central individuals in terrorist networks~\cite{LM4}. 

The analysis of \emph{evolving} social networks has long been regarded as a ``Holy Grail'' of network scientists~\cite{WF99,CSW7,SP0}. A number of models, mostly theoretical with limited empirical backing, have been considered for analyzing longitudinal social network data\cite{DS97,S7}, with networks that change either at discrete time points~\cite{KP59,WI88,BC96,RP1} or continuously over time~\cite{HL77,W80,S1}.  The complex network community has also been interested in the analysis of these dynamic networks; their research has mainly examined various mechanisms of network formation (\eg ``preferential attachment'' and its variants~\cite{AB2,DM2}), and the evolution of aggregate properties of the network at the macroscopic and mesoscopic levels (\eg network size, average degree, small-world characteristics, or community structure~\cite{AHKJ7,HEL4}).

While some of the above studies allow their networks to change, in most cases the evolution is assumed to happen more slowly than any dynamical processes occurring on the network.  Real network connections, however, are often more fluid than this; even when there exists an underlying fixed topological structure, connections between nodes can adaptively become active or inactive over time~\cite{SP0}. Consideration of the dynamical interplay between the state and the topology of the network is pertinent to a wide variety of network types. Examples include travel on a transportation network, message exchanges over communication networks, interactions between molecules that can bind to each other, and the switching of genes ``on'' and ``off'': all are networks with an underlying structure that can change quickly and adaptively with respect to the state of the network, and whose changes are relevant to the behavior and functionality of the system. The generality and broad implications of adaptive and dynamical wirings in real complex networks has been leading to a rapidly growing study and much interest from social scientists, physicists, and biologists (see reviews of recent advances in~\cite{GS9,GB8}). This mostly theoretical research has revealed a number of new approaches and insights including self-organization of adaptive networks~\cite{BR0,HG6,IK2}, contact processes and epidemiology on adaptive networks~\cite{GZ6,GDB6,SS8}, and social games on adaptive networks~\cite{SP0,ZES4,PTN6}.

Despite these important theoretical advances, there have been very few empirical studies related to the dynamics of large-scale network connections, or theoretical models that are validated and informed by empirical results~\cite{BB6,BB9,B5}.  Clearly, the study of data collected from real-world time-dependent networks, accompanied by theoretical models which address and reproduce the properties of said networks, will provide new, and perhaps fundamental insights into the mechanisms of dynamic networks.  For example, the experiments detailed in Ref.~\cite{BB6} detected a dramatic time dependence in network centrality and the role of nodes, something that is not apparent from static (\emph{i.e.,} time-aggregated) analysis of node connectivity and network topology. These experiments studied the time-dependent structure of a large-scale email network of over fifty-seven thousand users, based on data gathered over one hundred thirteen days from log files maintained by the email server at a large university. They found, among other things, that the daily networks (as well as the total aggregate network) were scale-free, but the ``hubs'' (that is, the most well-connected nodes) of these networks changed from day to day. In other words, a node which is among the most connected on one day, might only have a couple links the following day (popular today, anonymous tomorrow), and may not even be among the hubs of the aggregate network.  This phenomenon, called \textit{dynamic centrality}, is not accounted for by the standard explanation of hub formation: normally, hubs in networks develop in a system where popular nodes tend to attract more links (a process called \emph{preferential attachment}), so that when one particular node becomes slightly more popular than the rest, it attracts slightly more links from other nodes, making it more popular, and so forth in a process of positive feedback.  This process typically works over a long period of time, as news of a node's new ``popularity'' is allowed to disseminate through the system before the next connection is made.  In Ref.~\cite{BB6}, however, hubs develop over the course of a single day, and vanish just as quickly: too fast for such a process to occur.  Thus, the implications of dynamic centrality are in sharp contrast to previous complex network research, which depend on such preferential attachment-type mechanisms to create power-law distributions in empirical complex networks. This result draws attention to the inherent dynamic nature of adaptive networks which has long been neglected in the literature~\cite{GS9}. Similar results~\cite{BB9} were found in the interactions caused by the spatial proximity of personal Bluetooth wireless devices, recording the interactions between pairs of students over the period of 31 days. Results that are consistent with dynamic centrality behavior have also been seen in the social dynamics of free and open-source (FLOSS) development projects~\cite{WHC8}, peer prestige in academic hiring networks~\cite{WMA8}, trading networks~\cite{ABH9}, protein-interaction networks \cite{HBH4}, and transcriptional networks~\cite{LBY4}. The availability of complete (\emph{i.e.,} non-sampled) data that tracks the temporal dynamics of the interactions within the network will allow further study of dynamic centrality. 

The existence of dynamic centrality and a dynamic interaction structure has strong implications for existing work related to the attack and error tolerance of complex networks as well as in network transport. Research in epidemics has suggested that an effective disease or computer-virus prevention strategy would be to identify and vaccinate the high-degree nodes of a network, which would inhibit the spread of infection~\cite{DB2,PV2b,CHB3,GEM6}. Similarly, ``popular influencer'' marketing techniques (closely related to word-of-mouth or viral marketing) are based on the premise that focusing marketing activities on the hubs of social networks increases the likelihood of a cascading adoption of products or services, forming a type of social epidemic~\cite{R3}. The underlying assumption of both models is that the network topology is basically static: a node which is well-connected or popular now will continue to be a hub later in time.  In a dynamic network, however, well-connected nodes can quickly become only weakly connected (or even disconnected) tomorrow, which invalidates such strategies.  Thus, the existence of dynamic centrality calls for a radical rethinking of the interplay between link/structural dynamics and the dynamical processes underlying the time-dependent complex networks, and may call for modified strategies in predicting and preventing (or encouraging) epidemics.

Our paper contributes to the understanding of adaptive and dynamic networks by demonstrating a new approach to explain and reproduce real world dynamic centrality phenomena.  More specifically, we will investigate the type of network seen in Ref.~\cite{BB6, BB9}, which we will call \emph{dynamic scale-free} (DSF) networks; these are a series of scale-free networks with a scale-free aggregate and dynamic centrality.
We will show that DSF networks arise from a series of vertex-reinforced random walks on a finite scale-free network (called the \emph{underlay}), governed by a ``limited'' preferential-attachment reinforcement rule--- limited in the sense that, when making a connection, a node only takes into account the popularity of its neighbors in the underlay. 

The study of random walks on graphs, and reinforced random walks in particular, has a long fascinating history in its own right~\cite{DS84,P7}. In the context of social and complex networks, the nature of random walks and diffusion over small-world and scale-free networks is also a much investigated topic~\cite{JB0,JSB0,LKK1,MDL1,PA1,JB1,AKS2,LM3}. Random walks have been used in analyzing search and navigation on networks~\cite{ALP1}, calculating the ``betweenness centrality'' of nodes in a network~\cite{NG4}, extracting community structure from a network~\cite{NG4}, and sampling a large network as a means of estimating its topological properties~\cite{BP1,CT7}. The close interplay between the degree of a node and the proportion of time visited by a uniform random walk is a well-known result~\cite{DS84}, and has also been examined for various real world networks~\cite{CSA7}.  This body of literature largely focuses on the behavior of \emph{non-reinforced} random walks on static network topologies. In contrast to dynamics \emph{on} networks, a different approach for modeling the dynamics \emph{of} networks, which uses reinforced random walks or other stochastic processes that incorporate reinforcement mechanisms, has been considered~\cite{SP0,S7}. Most pertinent to our study are models of dynamic networks where individuals choose whom to interact with depending on how ``popular'' their choice is~\cite{SP0,W80}. Despite these theoretical advances, application to real world dynamic networks has been limited to parameter estimation with very small networks that are observed at several time points~\cite{S7}. Here we broaden the scope of this research in several distinctive and novel ways.  First, to the best of our knowledge, the use of a reinforced random walk has not yet been used as a mechanism of dynamical network connectivity that is informed and supported by actual high-frequency measurements of large-scale social network link dynamics. This shift in perspective leads to new insights including evidence that non-reinforced random walk schemes cannot serve as a plausible explanation for experimental data of dynamic networks, in contrast to reinforced random walk schemes. Second, random walk mechanisms (or other stochastic models) have not been used so far to study real-world dynamic centrality phenomena. Finally, new analysis methods, including various ``ranking'' and ``centrality overlap'' measures, are incorporated to facilitate and enable effective evaluation of models against large-scale data.

\section{Model}
In our approach we start with the aggregate network, which is called the \emph{underlay}; it is reasonable to assume that the aggregate network represents the long-term pattern of contact between individuals, which is often well approximated by a power-law distribution~\cite{AB2}.  Underlays are generated using the Barab\'asi-Albert algorithm \cite{BA99} with parameters $m_0=5$ (initial number of nodes; also the minimum degree of every node in the underlay) and  $N=1\,000\,000$ (the size of the network). We then generate a series of DSF networks using a vertex-reinforced random walk upon this underlay network: starting from a randomly chosen node, 
we add weight (according to a scheme described below) to the vertices every time they are visited, constantly altering the jump probabilities so that the walk will tend to revisit vertices already visited \cite{SP0, P7}. This proposal is inspired by the way messages on actual communication networks are part of larger-scale conversations. For instance, suppose Alice sends an email to Bob, who then responds to Alice and also forwards her message to Claire. Claire could in turn either forward it to someone new, reply to either Alice or Bob, or do nothing. Assuming these actions are roughly equal in likelihood, there is a greater probability that Claire sends an email to either Alice or Bob, than to some other specific individual; thus, people who are already part of the email chain are more likely to become part of the chain again. Here, each ``daily'' network is a subnetwork of the underlay, built on top of the underlay via the node-reinforced random walk mechanism.

The specific algorithm is as follows: starting with a randomly chosen node, we consider that node's neighbors $i$ in the underlay, each neighbor being weighted according to 
\begin{equation}
\label{weight}w_i=1+CV_i,
\end{equation}
where $V_i$ is the number of visits the node has received so far, and $C$ is a parameter of the model.  Neighbor $i$ is then chosen as a target with probability $\Pi_i=w_i/\sum_jw_j$.  Once a target is chosen, the link between the current node and the target is added to the sub-network, the algorithm steps to the target, and the process continues until the chain reaches the prescribed number of steps $S$.  The parameter $C$ captures the tendency of the random walk to revisit vertices--- that is, the strength of dynamic preferential attachment.  
When $C=0$, one has a pure (\emph{i.e.} unreinforced) random walk, a case we will refer to as a control, to demonstrate the importance of preferential attachment to our results.  It should be noted that the vertex-reinforced random walk is known~\cite{P7} to have significantly different behavior from the pure random walk.  For example, it is well-known that in a pure random walk the probability of visiting a particular node becomes proportional to the degree of the node, after $\log N$ steps.  This is not so for  vertex-reinforced random walks which can become ``trapped'' within a particular subnetwork\cite{P7}: nodes within the subnetwork are visited frequently while nodes outside the subnetwork, even high-degree nodes, remain unvisited.  This effect is critical to this model's ability to recreate dynamic centrality, but from a practical point of view it can be too powerful, trapping the walk in a very small subnetwork (even one of only two points).  To mollify this effect somewhat, we modify the vertex-reinforced rule to prohibit links from being traversed two times in a row: nodes and links can be revisited multiple times during the simulation, but a walk may not jump immediately backwards to the node it came from.

\section{Results \& Discussion}

\Inline{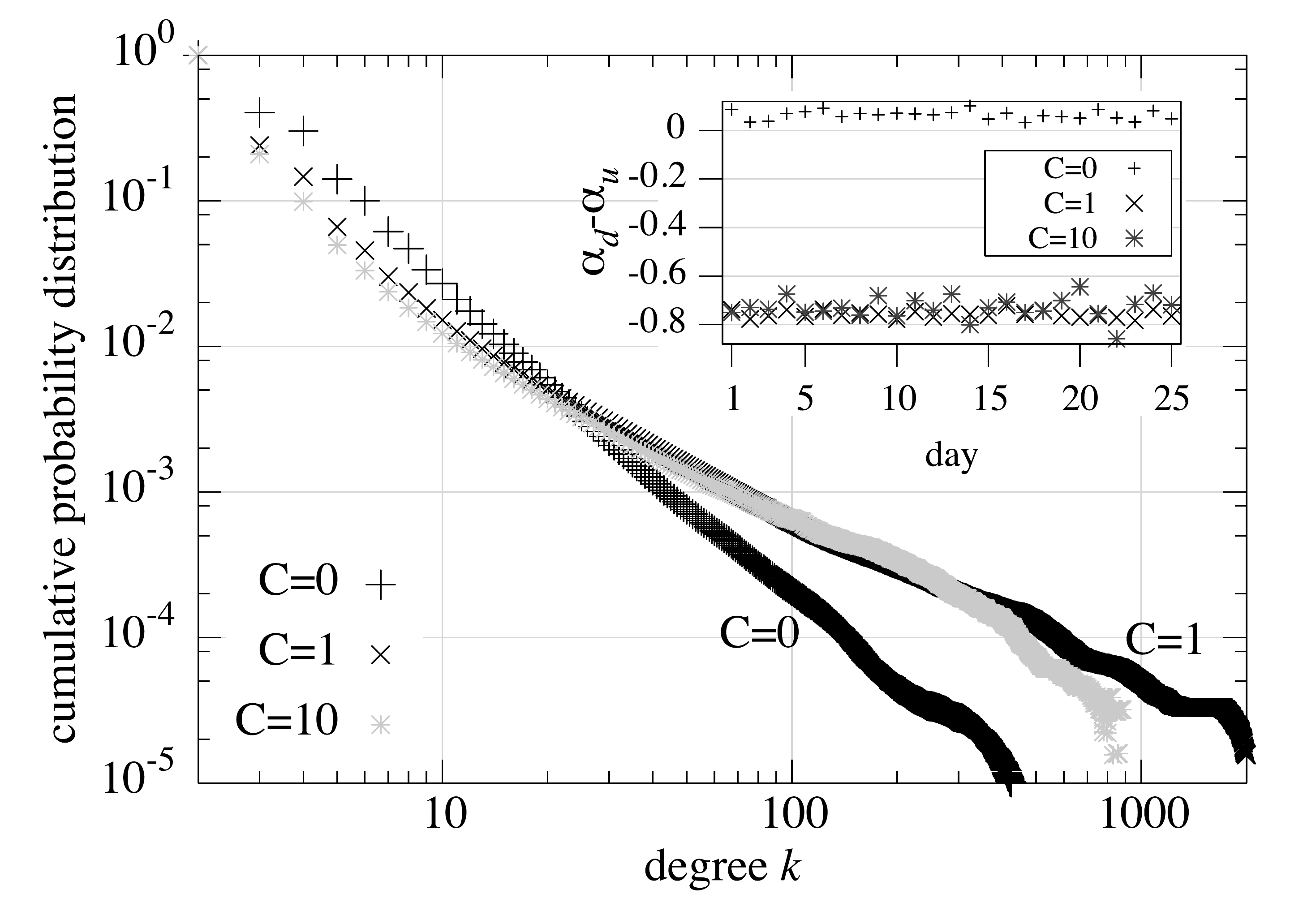}{histogram}{A log-log plot of the cumulative degree distributions of daily networks, averaged over 25 days, built with single chains of 1 million steps: each histogram shows the probability that a member node of the daily network has degree $k$ or larger.  All curves show power-law behavior; the inset shows the difference between the power-law exponent of each daily network's histogram, and the exponent of its underlay (approximately -1.9, though a different underlay was used for each value of {\it C}).} 

We will now show that the above reinforcement scheme is able to create scale-free networks with dynamic centrality and a scale-free aggregate (\emph{i.e.,} the underlay, scale-free by construction), reproducing the main qualitative features of DSF networks seen in~\cite{BB6,BB9}.  We begin by constructing daily subnetworks with $S$~=~1~million steps atop underlays of $N$~=~1~million nodes.  Figure~\ref{histogram} shows the cumulative probability degree distribution (averaged over 25 daily runs) for the cases of no ($C=0$), weak ($C=1$), or strong ($C=10$) reinforcement.  All three distributions show power-law behavior.  However, the networks with reinforcement ($C>0$) have smaller slopes and longer tails, which suggests a more heterogeneous degree distribution and a more pronounced scale-free behavior.  Furthermore, if we fit the histograms to a power-law $Ax^{-\alpha}$, the exponent $\alpha$ for $C=0$ is very close to the power-law exponent ($\alpha_u\approx2$) of the underlay\footnote{The underlay is a standard Barabasi-Albert network: the exponent is not the familiar $\alpha=3$ because we are dealing with the \emph{cumulative} degree distribution, which decreases the exponent by one.
}, suggesting that the random-walk network is merely taking on the scale-free nature of the underlay.  The inset shows this more clearly: considering 25 daily networks constructed for each value of $C$, we fit the cumulative histogram to a power-law, found the exponent $\alpha_d$ for each, and subtracted the exponent $\alpha_u$ of that network's underlay.  (In this case, we used three different underlays, one for each value of $C$.)  We see that the $C=0$ curve decays at the same rate as the underlay (or even a little more quickly); the difference is $\alpha_d-\alpha_u=0.06\pm0.02$. By contrast, the $C>0$ cases have $\alpha_d-\alpha_u=-0.76\pm0.01$ (for $C=1$) and $-0.73\pm0.04$ (for $C=10$).  With reinforcement, the daily exponents are smaller by about 0.75, resulting in a slower rate of decay and relatively more nodes of high degree.  Note also that these results are relatively consistent from day to day.  

Having shown that the subnetworks created by our model are scale-free subnetworks of a scale-free aggregate, we now demonstrate that they exhibit dynamic centrality.  Following \cite{BB6,BB9}, we define the ``Top--$n$ overlap'' for two subnetworks of a given underlay as follows: after ranking the nodes according to their degree in each subnetwork, we count the number of nodes which appear among the top $n$ nodes in both networks, dividing the count by $n$ to express it as a fraction (or percentage).  If the overlap is close to 100\%, then the two subnetworks' hubs are largely identical; thus we expect that daily subnetworks exhibiting dynamic centrality (where hubs change from day to day) have a relatively low Top--$n$ overlap, both with the underlay and with each other.  (The exact method and formula for calculating the overlap, which deals with the issue of ties for $n$th place, is described in the Appendix.)

\Inline{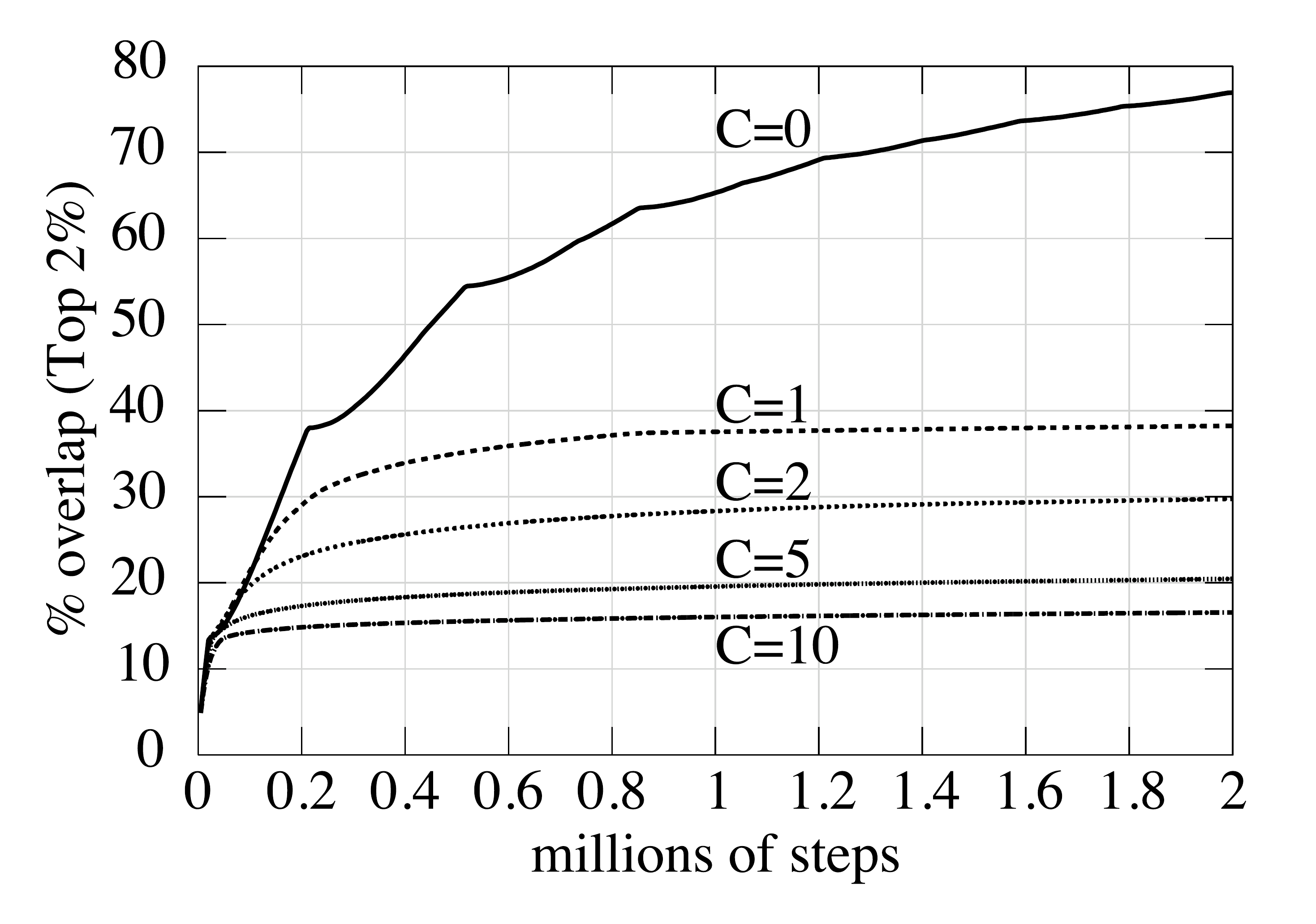}{overlap}{The Top--2\% overlap, as defined in the text, as a function of the number of steps $S$ in the daily network, for various values of $C$. The overlap for $C>0$ appears to increase at a much slower rate, suggesting that the highly connected nodes in the underlay are only modestly represented in the daily networks. }

We first used this ``top $n$ overlap'' to compare individual daily networks with their underlays.  In Fig.~\ref{overlap}, we see that the overlap of the top 2\% of nodes (\emph{i.e.,} the 20,000 nodes with the highest degree) is dramatically smaller with the introduction of dynamic preferential attachment, once the walk extends further than roughly 200,000 steps, and the overlap decreases monotonically as the preference weight $C$ increases.  While the overlap of the $C=0$ case grows steadily as the random walk fully explores the underlay, the overlap for $C>0$ increases at a much slower rate, suggesting that the highly connected nodes in the underlay are only modestly represented in the daily networks.

\Inline{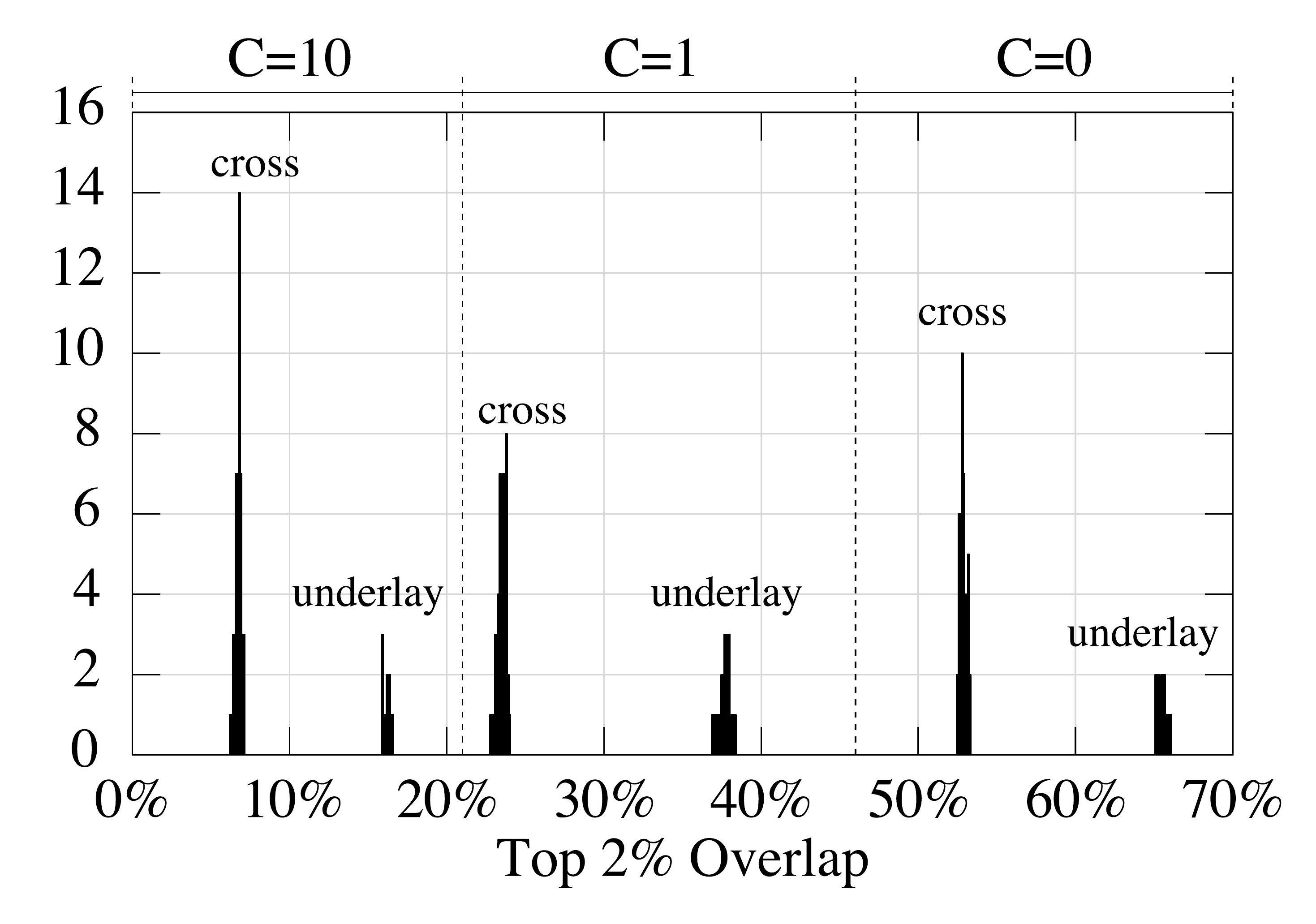}{dayvsday}{The distributions of the Top--2\% overlap between 10 daily networks and their common underlay (the peaks labelled ``underlay''), and between the 45 pairs of those 10 networks (labelled ``cross'').  The bar graphs were constructed by binning, with a bin size of $0.01\%$.  The results suggest that `local' hubs vary greatly from day to day.}

In Fig.~\ref{dayvsday}, we look at the distribution of the Top--2\% overlap between the underlay and 10 different daily subnetworks, generated with walks of $S=1$ million steps, and for $C=0$, $1$, and $10$; we also look at the Top--2\% overlap for the 45 pairs one can make among the 10 daily networks for each value of $C$.  All the distributions are sharply peaked, showing that the overlap does not vary much from day to day, and the overlap is again greatly decreased by the introduction of reinforcement, even from day to day.  This demonstrates that the set of hubs of the daily network varies greatly from day to day, a result consistent with empirical dynamic scale-free networks~\cite{BB6,BB9}.

\Inline{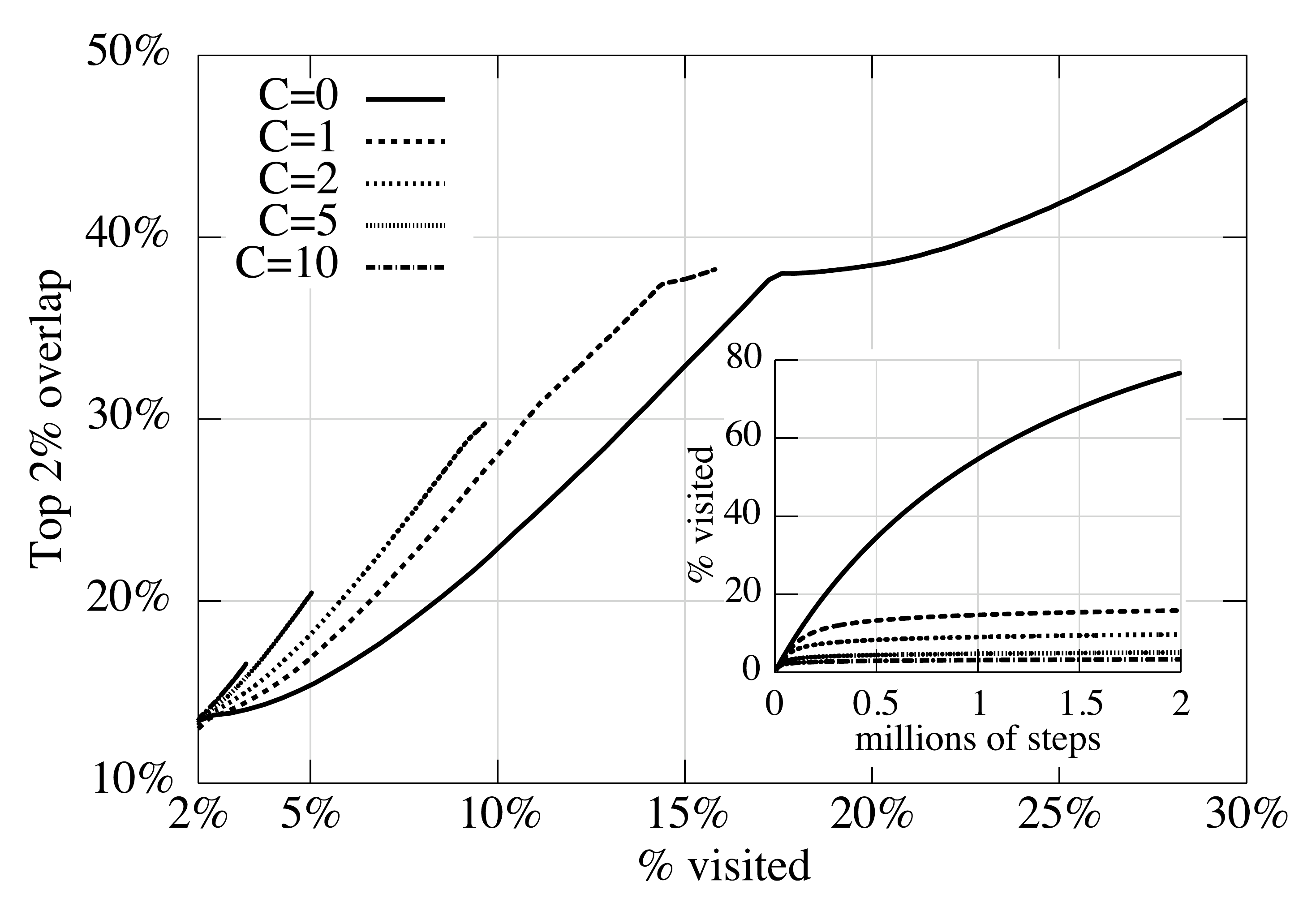}{active}{The importance of the number of unique visited nodes.  The inset shows the fraction of nodes in the underlay that are included in a daily network generated with different numbers of steps $S$ and with $C=0,1,2,5,$ and $10$. Higher percentages of unique visited nodes correspond monotonically to lower values of $C$. The main graph shows the same curves plotted versus the Top--2\% overlap instead of the number of steps.  The kink in the $C=0$ curve is a finite-size effect. 
}

An explanation for dynamic centrality in the $C>0$ case can be found by examining the number of unique nodes which are visited by the random walk, which is equivalent to the number of nodes in the resulting daily network, as is done in Fig.~\ref{active}.  For the pure random walk ($C=0$), the underlay network (due to its small-world nature) is rapidly covered as the number of steps $S$ increases, as is seen in the inset.  The Top--2\% overlap (shown in the main portion of the figure) is closely correlated to the number of nodes in the daily network, so as the daily network grows, it increasingly adopts the underlay's hubs as its own, ruining dynamic centrality because the hubs remain constant from day to day.  With reinforcement ($C>0$), however, the random walk is more likely to revisit nodes it has visited before, and so it takes much longer to explore the underlay. This slow rate of exploration appears to be the primary reason that the overlap is suppressed in Fig.~\ref{overlap}; in fact, Fig.~\ref{active} shows that if we compare daily networks with the same number of \emph{nodes}, rather than the same number of steps, reinforcement actually results in a slightly \emph{higher} overlap.

\Inline{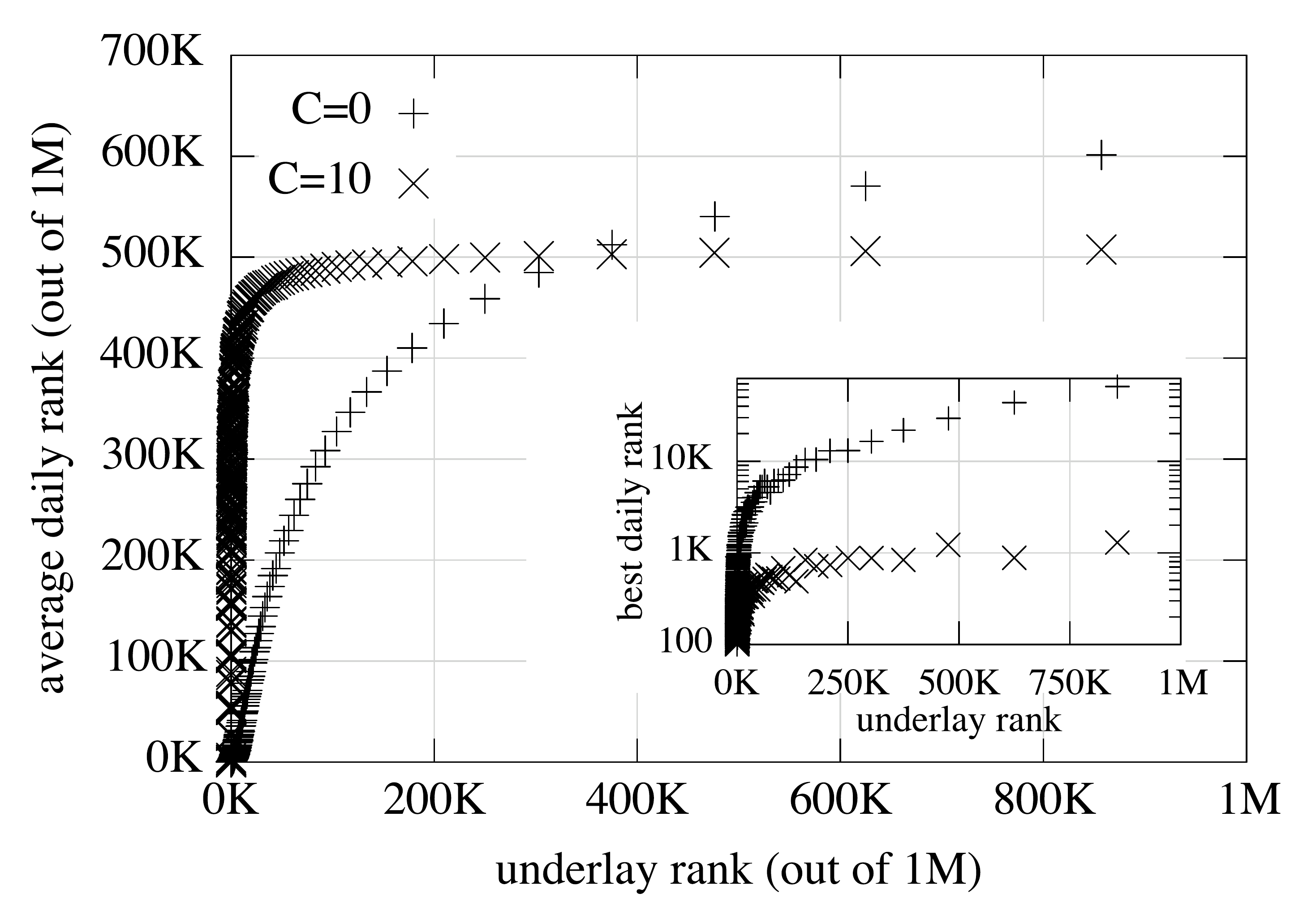}{rank}{Comparison of the underlay network with daily networks.  We show the average rank of a node in 10 daily subnetworks (with 1 million steps) versus their rank in the underlay network; nodes with higher degree have lower rank, and the top-ranked node has rank 0.  In the case of ties, all tied nodes are assigned the rank that is the average value of the range they cover.  The inset shows, not the average, but the best daily rank that node has over the 10 days.  
The $C>0$ case is remarkably egalitarian - every node has an opportunity to become a hub in some daily network.
}
We have so far looked at the behavior of the top 20,000 nodes in the underlay and their modest role  in the daily networks, but we can also compare the rankings of \emph{all} nodes in the underlay with their ranking in the daily subnetworks.  We did so as follows: for a given underlay, we ranked each node according to its degree, with the most highly connected node being given rank 0; nodes with the same degree were grouped together and given a rank which was the average of the range they covered.  We then generated 10 daily subnetworks of $S$=1 million steps each, and averaged the rank each node had over all ten subnetworks, dealing with ties in the same way as above.  We did this for $C=0$ and $C=10$, and Fig.~\ref{rank} shows the underlay rank plotted versus the average daily rank so calculated.  When multiple nodes have the same underlay rank, the average daily rank has itself been averaged over \emph{all} nodes with the same underlay rank.  In the $C=0$ case, we see that the underlay and daily ranks have a strong positive correlation: the daily rank of a node always depends strongly on its rank in the underlay.  The $C=10$ curve is much different: the average daily rank is roughly \emph{constant}, once we consider nodes below the top 50,000 (5\%) or so.  In addition to the average daily rank, we can also consider the \emph{best} (i.e. minimum) daily rank that a node reaches over the 10 daily networks; in doing so, we see (in the inset to Fig.~\ref{rank}) that \emph{every} node in the $C>0$ case has an opportunity to be among the top 1500 nodes (0.15\%) in \emph{some} daily network; without preferential attachment, most nodes can't crack the top 10,000 (1\%).  The $C>0$ case is extremely \emph{egalitarian} in this regard: any node, even those with the smallest degree in the underlay, can be among the most important nodes in some daily network.   This behavior agrees with the empirical results reported in \cite{BB6,BB9}.

\Inline{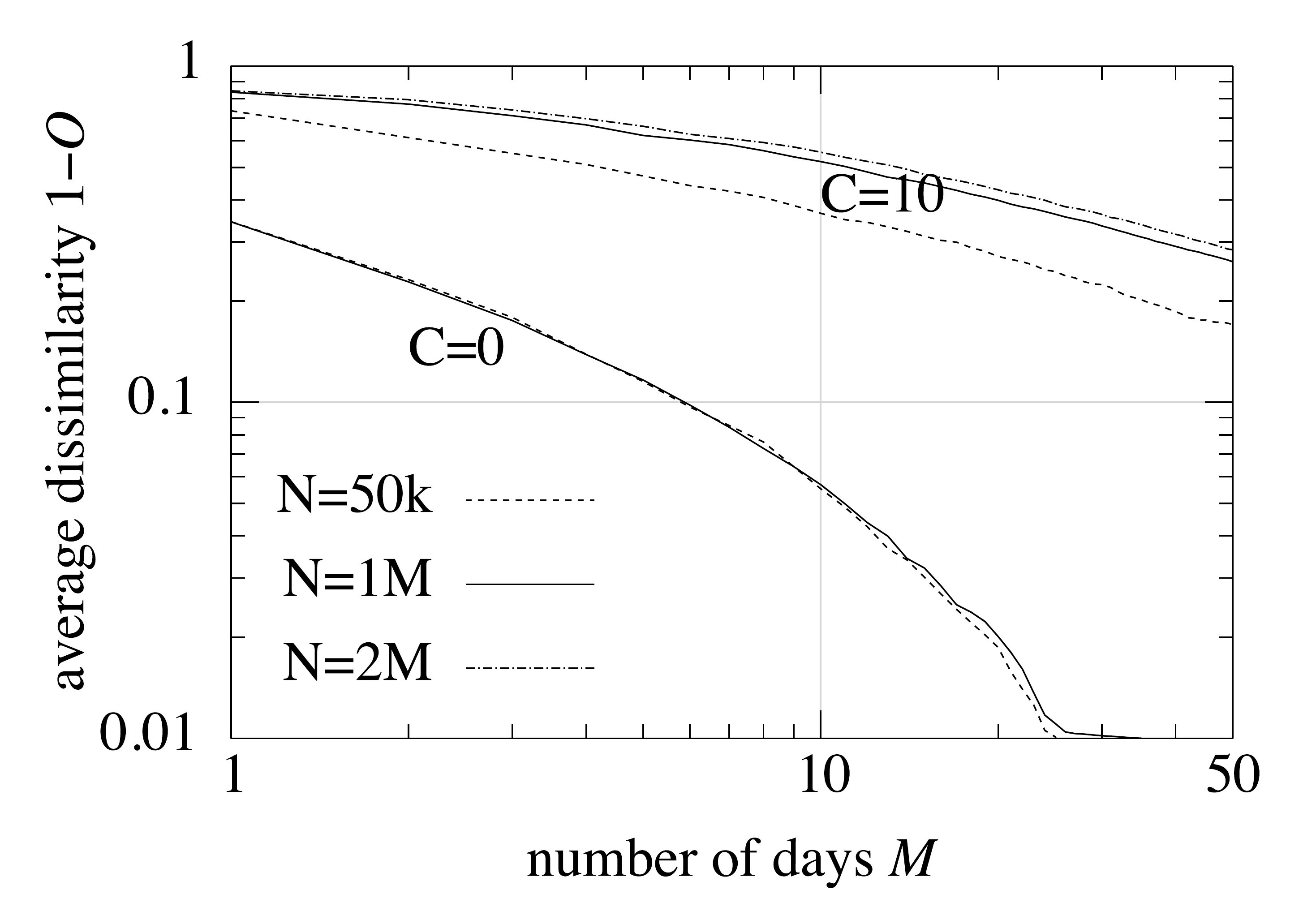}{multiple}{Dissimilarity of networks aggregated over times ranging from 1 to 50 days.  Dissimilarity is defined as the Top--2\% overlap $O$ (Eq.~2) subtracted from 100\%, and is shown for both pure random walks ($C=0$) and strongly weighted walks ($C=10$).  We show results for networks with 50 thousand, 1 million, and 2 million nodes to show how the system scales; all walks have as many steps as there are nodes ($S=N$).  
}


One last empirical result seen in \cite{BB6} involves the aggregate of $M$ daily subnetworks, which are formed by connecting two nodes together if they are connected in any of the $M$ subnetworks.
Given their independence and random starting points, the aggregate of $M$ daily networks must eventually converge to the underlay as $M$ grows large, but whether this convergence occurs over a particular timescale $M_0$, or whether it converges in a scale-free manner, is an interesting question.  One way to measure this convergence is by calculating the Top--2\% overlap ${\cal O}$ between the aggregate and the underlay, and subtracting this from 100\%; if this quantity (called the ``dissimilarity'' reaches zero, then both the aggregate and the underlay both have the same set of hubs, and are roughly (though not precisely) equivalent.  Fig.~\ref{multiple} shows this dissimilarity as a function of the aggregate size $M$.  For the pure random walk case ($C=0$), the dissimilarity decreases exponentially ($\sim e^{-M/M_0}$, with $M_0\approx -7.5$), so that after 20 days the top nodes in both networks are basically identical.  In contrast, the aggregate generated with a strong dynamic preferential attachment ($C=10$) has a dissimilarity which decays much more slowly, and possibly without a characteristic timescale.  This is the behavior seen in empirical DSF networks~\cite{BB6,BB9}, and hints at a ``multiscale'' structure for the network dynamics, where networks at each timescale form scale-free topologies while the specific links in existence vary dramatically between observation timescales as well as over time.  To alleviate concerns that this disparity may be a finite-size effect, we include data in Fig.~\ref{multiple} for networks of $N=50\,000$ nodes and $N=2\,000\,000$ nodes (where $S=N$ in all cases): increasing the network size does nothing to change the $C=0$ case, and actually results in an even slower rate of decay for the reinforced $C=10$ networks.

\section{Conclusion}
We have embarked on a research program designed to develop universal models that can recreate empirically observed phenomena in dynamic complex networks.  We have shown that, using a suitable reinforced random walk on a ``long-term'' underlay network, one is able to produce instantaneous networks which reproduce qualitatively characteristic features of real world dynamic networks. This includes, in particular, the construction of scale-free sub-networks of a scale-free ``underlay'' network, whose local hubs substantially differ from sub-network to sub-network and from those of the underlay. We have presented evidence that dynamic preferential attachment (as opposed to a pure random walk) is a necessary requirement of dynamic scale-free networks: the preferential attachment mechanism limits the walk's exploration of the network, which gives early-visited nodes higher ranks than they would have with a more random traversal. We hope our model will stimulate further empirical and theoretical work, and provide a framework for analyzing the influence of link/structural dynamics on dynamical processes on complex networks.

\appendix
\section{Calculating the top--\emph{n} overlap}

If there are no ties for $n$th place in the network, then the top--$n$ overlap of subnetworks ${\cal A}$ and ${\cal B}$ is given by the formula
\begin{equation}
{\cal O}=\frac{1}{n}\abs{A\top\cap B\top}\label{eq-noties},
\end{equation}
where $A\top$ is a set of the top $n$ nodes in network ${\cal A}$, as $B\top$ is for ${\cal B}$, and $\abs{\cdot}$ denotes the number of elements in a set.  Given the finite size of our networks, however, there are generally large groups of nodes with the same degree, making the choice of the top $n$ nodes somewhat arbitrary.  To account for this, we modify Eq.~\ref{eq-noties} to prorate those nodes which are tied for the $n$th slot, as follows:  let $a$ and $b$ be the degree of the $n$th highest node in networks $\cal A$ and $\cal B$, respectively.  We define the set $A_>$ to be the set of nodes in $\cal A$ with degree greater than $a$, while $A_=$ is the (guaranteed nonempty) set of nodes in ${\cal A}$ with degree equal to $a$; we define $B_>$ and $B_=$ similarly.  Imagine constructing the list $A\top$ of the top $n$ nodes: it will automatically contain the entire set $A_>$, and will have room for $n-\abs{A_>}$ of the nodes in $A_=$.  If we choose nodes at random from $A_=$ to fill those available slots, then each node from $A_=$ has probability
\begin{equation}
f_A={n-\abs{A_>}\over \abs{A_=}}\label{eq-fraction}.
\end{equation}
of appearing in the list $A\top$.  (The same applies for network ${\cal B}$.)
Thus, when calculating the overlap between ${\cal A}$ and ${\cal B}$, we will count each node in $A_=$ and $B_=$ as being worth a fraction of a regular node--- $f_A$ and $f_B$, respectively--- define the modified overlap $O({\cal A},{\cal B})$ to be
\begin{eqnarray}
{\cal O}&=&\frac{1}{n}\Bigl[\abs{A_>\cap B_>}+f_A\abs{A_=\cap B_>}\cr&&
+f_B\abs{A_>\cap B_=}+f_Af_B\abs{A_=\cap B_=}\Bigr]\label{eq-overlap}
\end{eqnarray}

Note that if there is no tie for $n$th place in either network, then $f_A=f_B=1$, $A\top=(A_>\cup A_=)$, and $B\top=(B_>\cup B_=)$; and Eq.~\ref{eq-overlap} reduces to the simpler Eq.~\ref{eq-noties} described above.

%

\Figure{Fig1.pdf}{histogram}{A log-log plot of the cumulative degree distributions of daily networks, averaged over 25 days, built with single chains of 1 million steps: each histogram shows the probability that a member node of the daily network has degree $k$ or larger.  All curves show power-law behavior; the inset shows the difference between the power-law exponent of each daily network's histogram, and the exponent of its underlay (approximately -1.9, though a different underlay was used for each value of {\it C}).} 

\Figure{Fig2.pdf}{overlap}{The Top--2\% overlap, as defined in the text, as a function of the number of steps $S$ in the daily network, for various values of $C$. The overlap for $C>0$ appears to increase at a much slower rate, suggesting that the highly connected nodes in the underlay are only modestly represented in the daily networks. }
\Figure{Fig3.pdf}{dayvsday}{The distributions of the Top--2\% overlap between 10 daily networks and their common underlay (the peaks labelled ``underlay''), and between the 45 pairs of those 10 networks (labelled ``cross'').  The bar graphs were constructed by binning, with a bin size of $0.01\%$.  The results suggest that `local' hubs vary greatly from day to day.}
\Figure{Fig4.pdf}{active}{The importance of the number of unique visited nodes.  The inset shows the fraction of nodes in the underlay that are included in a daily network generated with different numbers of steps $S$ and with $C=0,1,2,5,$ and $10$. Higher percentages of unique visited nodes correspond monotonically to lower values of $C$. The main graph shows the same curves plotted versus the Top--2\% overlap instead of the number of steps.  The kink in the $C=0$ curve is a finite-size effect. 
}
\Figure{Fig5.pdf}{rank}{Comparison of the underlay network with daily networks.  We show the average rank of a node in 10 daily subnetworks (with 1 million steps) versus their rank in the underlay network; nodes with higher degree have lower rank, and the top-ranked node has rank 0.  In the case of ties, all tied nodes are assigned the rank that is the average value of the range they cover.  The inset shows, not the average, but the best daily rank that node has over the 10 days.  
The $C>0$ case is remarkably egalitarian - every node has an opportunity to become a hub in some daily network.
}
\Figure{Fig6.pdf}{multiple}{Dissimilarity of networks aggregated over times ranging from 1 to 50 days.  Dissimilarity is defined as the Top--2\% overlap $O$ (Eq.~2) subtracted from 100\%, and is shown for both pure random walks ($C=0$) and strongly weighted walks ($C=10$).  We show results for networks with 50 thousand, 1 million, and 2 million nodes to show how the system scales; all walks have as many steps as there are nodes ($S=N$).  
}



\begin{thebibliography}{65}%
\makeatletter
\providecommand \@ifxundefined [1]{%
 \@ifx{#1\undefined}
}%
\providecommand \@ifnum [1]{%
 \ifnum #1\expandafter \@firstoftwo
 \else \expandafter \@secondoftwo
 \fi
}%
\providecommand \@ifx [1]{%
 \ifx #1\expandafter \@firstoftwo
 \else \expandafter \@secondoftwo
 \fi
}%
\providecommand \@bibitemShut {\relax}
\providecommand \natexlab [1]{#1}%
\providecommand \enquote  [1]{``#1''}%
\providecommand \bibnamefont  [1]{#1}%
\providecommand \bibfnamefont [1]{#1}%
\providecommand \citenamefont [1]{#1}%
\providecommand \href@noop [0]{\@secondoftwo}%
\providecommand \href [0]{\begingroup \@sanitize@url \@href}%
\providecommand \@href[1]{\@@startlink{#1}\@@href}%
\providecommand \@@href[1]{\endgroup#1\@@endlink}%
\providecommand \@sanitize@url [0]{\catcode `\\12\catcode `\$12\catcode
  `\&12\catcode `\#12\catcode `\^12\catcode `\_12\catcode `\%12\relax}%
\providecommand \@@startlink[1]{}%
\providecommand \@@endlink[0]{}%
\providecommand \url  [0]{\begingroup\@sanitize@url \@url }%
\providecommand \@url [1]{\endgroup\@href {#1}{\urlprefix }}%
\providecommand \urlprefix  [0]{URL }%
\providecommand \Eprint [0]{\href }%
\@ifxundefined \urlstyle {%
  \providecommand \doi  [0]{\begingroup \@sanitize@url \@doi}%
  \providecommand \@doi [1]{\endgroup \@@startlink {\doibase
  #1}doi:\discretionary {}{}{}#1\@@endlink }%
}{%
  \providecommand \doi  [0]{doi:\discretionary{}{}{}\begingroup
  \urlstyle{rm}\Url }%
}%
\providecommand \doibase [0]{http://dx.doi.org/}%
\providecommand \Doi [0]{\begingroup \@sanitize@url \@Doi }%
\providecommand \@Doi  [1]{\endgroup\@@startlink{\doibase#1}\@@Doi}%
\providecommand \@@Doi [1]{#1\@@endlink}%
\providecommand \selectlanguage [0]{\@gobble}%
\providecommand \bibinfo  [0]{\@secondoftwo}%
\providecommand \bibfield  [0]{\@secondoftwo}%
\providecommand \translation [1]{[#1]}%
\providecommand \BibitemOpen [0]{}%
\providecommand \bibitemStop [0]{}%
\providecommand \bibitemNoStop [0]{.\EOS\space}%
\providecommand \EOS [0]{\spacefactor3000\relax}%
\providecommand \BibitemShut  [1]{\csname bibitem#1\endcsname}%
\bibitem [{\citenamefont {Wasserman}\ and\ \citenamefont {Faust}(1999)}]{WF99}%
  \BibitemOpen
  \bibfield  {author} {\bibinfo {author} {\bibfnamefont {S.}~\bibnamefont
  {Wasserman}}\ and\ \bibinfo {author} {\bibfnamefont {K.}~\bibnamefont
  {Faust}},\ }\href@noop {} {\emph {\bibinfo {title} {Social Network
  Analysis}}}\ (\bibinfo  {publisher} {Cambridge University Press},\ \bibinfo
  {address} {Cambridge, UK},\ \bibinfo {year} {1999})\BibitemShut {NoStop}%
\bibitem [{\citenamefont {Carrington}\ \emph {et~al.}(2007)\citenamefont
  {Carrington}, \citenamefont {Scott},\ and\ \citenamefont {Wasserman}}]{CSW7}%
  \BibitemOpen
  \bibinfo {editor} {\bibfnamefont {P.~J.}\ \bibnamefont {Carrington}},
  \bibinfo {editor} {\bibfnamefont {J.}~\bibnamefont {Scott}}, \ and\ \bibinfo
  {editor} {\bibfnamefont {S.}~\bibnamefont {Wasserman}},\ eds.,\ \href@noop {}
  {\emph {\bibinfo {title} {Models and Methods in Social Network Analysis}}}\
  (\bibinfo  {publisher} {Cambridge University Press},\ \bibinfo {address}
  {Cambridge, UK},\ \bibinfo {year} {2007})\BibitemShut {NoStop}%
\bibitem [{\citenamefont {Albert}\ and\ \citenamefont
  {Barab{\'a}si}(2002)}]{AB2}%
  \BibitemOpen
  \bibfield  {author} {\bibinfo {author} {\bibfnamefont {R.}~\bibnamefont
  {Albert}}\ and\ \bibinfo {author} {\bibfnamefont {A.-L.}\ \bibnamefont
  {Barab{\'a}si}},\ }\href@noop {} {\bibfield  {journal} {\bibinfo  {journal}
  {Rev. Mod. Phys.},\ }\textbf {\bibinfo {volume} {74}},\ \bibinfo {pages} {47}
  (\bibinfo {year} {2002})}\BibitemShut {NoStop}%
\bibitem [{\citenamefont {Amaral}\ \emph {et~al.}(2000)\citenamefont {Amaral},
  \citenamefont {Scala}, \citenamefont {Barth{\'e}l{\'e}my},\ and\
  \citenamefont {Stanley}}]{ASB0}%
  \BibitemOpen
  \bibfield  {author} {\bibinfo {author} {\bibfnamefont {L.~A.~N.}\
  \bibnamefont {Amaral}}, \bibinfo {author} {\bibfnamefont {A.}~\bibnamefont
  {Scala}}, \bibinfo {author} {\bibfnamefont {M.}~\bibnamefont
  {Barth{\'e}l{\'e}my}}, \ and\ \bibinfo {author} {\bibfnamefont {H.~E.}\
  \bibnamefont {Stanley}},\ }\href@noop {} {\bibfield  {journal} {\bibinfo
  {journal} {Proc. Nat. Acad. Sci. USA},\ }\textbf {\bibinfo {volume} {97}},\
  \bibinfo {pages} {11149} (\bibinfo {year} {2000})}\BibitemShut {NoStop}%
\bibitem [{\citenamefont {Boccaletti}\ \emph {et~al.}(2006)\citenamefont
  {Boccaletti}, \citenamefont {Latora}, \citenamefont {Moreno}, \citenamefont
  {Chaves},\ and\ \citenamefont {Hwang}}]{BLM6}%
  \BibitemOpen
  \bibfield  {author} {\bibinfo {author} {\bibfnamefont {S.}~\bibnamefont
  {Boccaletti}}, \bibinfo {author} {\bibfnamefont {V.}~\bibnamefont {Latora}},
  \bibinfo {author} {\bibfnamefont {Y.}~\bibnamefont {Moreno}}, \bibinfo
  {author} {\bibfnamefont {M.}~\bibnamefont {Chaves}}, \ and\ \bibinfo {author}
  {\bibfnamefont {D.-U.}\ \bibnamefont {Hwang}},\ }\href@noop {} {\bibfield
  {journal} {\bibinfo  {journal} {Physics Reports},\ }\textbf {\bibinfo
  {volume} {424}},\ \bibinfo {pages} {175} (\bibinfo {year}
  {2006})}\BibitemShut {NoStop}%
\bibitem [{\citenamefont {da~Fontoura~Costa}\ \emph
  {et~al.}(2007){\natexlab{a}}\citenamefont {da~Fontoura~Costa}, \citenamefont
  {Rodrigues}, \citenamefont {Travieso},\ and\ \citenamefont {Boas}}]{CRT7}%
  \BibitemOpen
  \bibfield  {author} {\bibinfo {author} {\bibfnamefont {L.}~\bibnamefont
  {da~Fontoura~Costa}}, \bibinfo {author} {\bibfnamefont {F.~A.}\ \bibnamefont
  {Rodrigues}}, \bibinfo {author} {\bibfnamefont {G.}~\bibnamefont {Travieso}},
  \ and\ \bibinfo {author} {\bibfnamefont {P.~R.~V.}\ \bibnamefont {Boas}},\
  }\href@noop {} {\bibfield  {journal} {\bibinfo  {journal} {Adv. Phys.},\
  }\textbf {\bibinfo {volume} {56}},\ \bibinfo {pages} {167} (\bibinfo {year}
  {2007}{\natexlab{a}})}\BibitemShut {NoStop}%
\bibitem [{\citenamefont {Barrat}\ \emph {et~al.}(2008)\citenamefont {Barrat},
  \citenamefont {Barth{\'e}l{\'e}my},\ and\ \citenamefont {Vespignani}}]{BBV8}%
  \BibitemOpen
  \bibfield  {author} {\bibinfo {author} {\bibfnamefont {A.}~\bibnamefont
  {Barrat}}, \bibinfo {author} {\bibfnamefont {M.}~\bibnamefont
  {Barth{\'e}l{\'e}my}}, \ and\ \bibinfo {author} {\bibfnamefont
  {A.}~\bibnamefont {Vespignani}},\ }\href@noop {} {\emph {\bibinfo {title}
  {Dynamical Processes on Complex Networks}}}\ (\bibinfo  {publisher}
  {Cambridge University Press},\ \bibinfo {address} {Cambridge, UK},\ \bibinfo
  {year} {2008})\BibitemShut {NoStop}%
\bibitem [{\citenamefont {Barab{\'a}si}\ and\ \citenamefont
  {Albert}(1999)}]{BA99}%
  \BibitemOpen
  \bibfield  {author} {\bibinfo {author} {\bibfnamefont {A.-L.}\ \bibnamefont
  {Barab{\'a}si}}\ and\ \bibinfo {author} {\bibfnamefont {R.}~\bibnamefont
  {Albert}},\ }\href@noop {} {\bibfield  {journal} {\bibinfo  {journal}
  {Science},\ }\textbf {\bibinfo {volume} {286}},\ \bibinfo {pages} {509}
  (\bibinfo {year} {1999})}\BibitemShut {NoStop}%
\bibitem [{\citenamefont {Dorogovtsev}\ and\ \citenamefont
  {Mendes}(2002)}]{DM2}%
  \BibitemOpen
  \bibfield  {author} {\bibinfo {author} {\bibfnamefont {S.~N.}\ \bibnamefont
  {Dorogovtsev}}\ and\ \bibinfo {author} {\bibfnamefont {J.~F.~F.}\
  \bibnamefont {Mendes}},\ }\href@noop {} {\bibfield  {journal} {\bibinfo
  {journal} {Adv. Phys.},\ }\textbf {\bibinfo {volume} {51}},\ \bibinfo {pages}
  {1079} (\bibinfo {year} {2002})}\BibitemShut {NoStop}%
\bibitem [{\citenamefont {Albert}\ \emph {et~al.}(2000)\citenamefont {Albert},
  \citenamefont {Jeong},\ and\ \citenamefont {Barab{\'a}si}}]{AJB0}%
  \BibitemOpen
  \bibfield  {author} {\bibinfo {author} {\bibfnamefont {R.}~\bibnamefont
  {Albert}}, \bibinfo {author} {\bibfnamefont {H.}~\bibnamefont {Jeong}}, \
  and\ \bibinfo {author} {\bibfnamefont {A.-L.}\ \bibnamefont {Barab{\'a}si}},\
  }\href@noop {} {\bibfield  {journal} {\bibinfo  {journal} {Nature},\ }\textbf
  {\bibinfo {volume} {406}},\ \bibinfo {pages} {378} (\bibinfo {year}
  {2000})}\BibitemShut {NoStop}%
\bibitem [{\citenamefont {Cohen}\ \emph {et~al.}(2000)\citenamefont {Cohen},
  \citenamefont {Erez}, \citenamefont {ben Avraham},\ and\ \citenamefont
  {Havlin}}]{CEB0}%
  \BibitemOpen
  \bibfield  {author} {\bibinfo {author} {\bibfnamefont {R.}~\bibnamefont
  {Cohen}}, \bibinfo {author} {\bibfnamefont {K.}~\bibnamefont {Erez}},
  \bibinfo {author} {\bibfnamefont {D.}~\bibnamefont {ben Avraham}}, \ and\
  \bibinfo {author} {\bibfnamefont {S.}~\bibnamefont {Havlin}},\ }\href@noop {}
  {\bibfield  {journal} {\bibinfo  {journal} {Phys. Rev. Lett.},\ }\textbf
  {\bibinfo {volume} {85}},\ \bibinfo {pages} {4626} (\bibinfo {year}
  {2000})}\BibitemShut {NoStop}%
\bibitem [{\citenamefont {Cohen}\ \emph {et~al.}(2001)\citenamefont {Cohen},
  \citenamefont {Erez}, \citenamefont {ben Avraham},\ and\ \citenamefont
  {Havlin}}]{CEB1}%
  \BibitemOpen
  \bibfield  {author} {\bibinfo {author} {\bibfnamefont {R.}~\bibnamefont
  {Cohen}}, \bibinfo {author} {\bibfnamefont {K.}~\bibnamefont {Erez}},
  \bibinfo {author} {\bibfnamefont {D.}~\bibnamefont {ben Avraham}}, \ and\
  \bibinfo {author} {\bibfnamefont {S.}~\bibnamefont {Havlin}},\ }\href@noop {}
  {\bibfield  {journal} {\bibinfo  {journal} {Phys. Rev. Lett.},\ }\textbf
  {\bibinfo {volume} {86}},\ \bibinfo {pages} {3682} (\bibinfo {year}
  {2001})}\BibitemShut {NoStop}%
\bibitem [{\citenamefont {Pastor-Satorras}\ and\ \citenamefont
  {Vespignani}(2002){\natexlab{a}}}]{PV2a}%
  \BibitemOpen
  \bibfield  {author} {\bibinfo {author} {\bibfnamefont {R.}~\bibnamefont
  {Pastor-Satorras}}\ and\ \bibinfo {author} {\bibfnamefont {A.}~\bibnamefont
  {Vespignani}},\ }\href@noop {} {\bibfield  {journal} {\bibinfo  {journal}
  {Phys. Rev. E},\ }\textbf {\bibinfo {volume} {65}},\ \bibinfo {pages}
  {035108} (\bibinfo {year} {2002}{\natexlab{a}})}\BibitemShut {NoStop}%
\bibitem [{\citenamefont {Desz{\"o}}\ and\ \citenamefont
  {Barab{\'a}si}(2002)}]{DB2}%
  \BibitemOpen
  \bibfield  {author} {\bibinfo {author} {\bibfnamefont {Z.}~\bibnamefont
  {Desz{\"o}}}\ and\ \bibinfo {author} {\bibfnamefont {A.-L.}\ \bibnamefont
  {Barab{\'a}si}},\ }\href@noop {} {\bibfield  {journal} {\bibinfo  {journal}
  {Phys. Rev. E},\ }\textbf {\bibinfo {volume} {65}},\ \bibinfo {pages}
  {055103} (\bibinfo {year} {2002})}\BibitemShut {NoStop}%
\bibitem [{\citenamefont {Pastor-Satorras}\ and\ \citenamefont
  {Vespignani}(2002){\natexlab{b}}}]{PV2b}%
  \BibitemOpen
  \bibfield  {author} {\bibinfo {author} {\bibfnamefont {R.}~\bibnamefont
  {Pastor-Satorras}}\ and\ \bibinfo {author} {\bibfnamefont {A.}~\bibnamefont
  {Vespignani}},\ }\href@noop {} {\bibfield  {journal} {\bibinfo  {journal}
  {Phys. Rev. E},\ }\textbf {\bibinfo {volume} {65}},\ \bibinfo {pages}
  {036104} (\bibinfo {year} {2002}{\natexlab{b}})}\BibitemShut {NoStop}%
\bibitem [{\citenamefont {Cohen}\ \emph {et~al.}(2003)\citenamefont {Cohen},
  \citenamefont {Havlin},\ and\ \citenamefont {ben Avraham}}]{CHB3}%
  \BibitemOpen
  \bibfield  {author} {\bibinfo {author} {\bibfnamefont {R.}~\bibnamefont
  {Cohen}}, \bibinfo {author} {\bibfnamefont {S.}~\bibnamefont {Havlin}}, \
  and\ \bibinfo {author} {\bibfnamefont {D.}~\bibnamefont {ben Avraham}},\
  }\href@noop {} {\bibfield  {journal} {\bibinfo  {journal} {Phys. Rev.
  Lett.},\ }\textbf {\bibinfo {volume} {91}},\ \bibinfo {pages} {247901}
  (\bibinfo {year} {2003})}\BibitemShut {NoStop}%
\bibitem [{\citenamefont {G{\'o}mez-Garde{\~n}es}\ \emph
  {et~al.}(2006)\citenamefont {G{\'o}mez-Garde{\~n}es}, \citenamefont
  {Echenique},\ and\ \citenamefont {Moreno}}]{GEM6}%
  \BibitemOpen
  \bibfield  {author} {\bibinfo {author} {\bibfnamefont {J.}~\bibnamefont
  {G{\'o}mez-Garde{\~n}es}}, \bibinfo {author} {\bibfnamefont {P.}~\bibnamefont
  {Echenique}}, \ and\ \bibinfo {author} {\bibfnamefont {Y.}~\bibnamefont
  {Moreno}},\ }\href@noop {} {\bibfield  {journal} {\bibinfo  {journal} {The
  European Physical Journal B},\ }\textbf {\bibinfo {volume} {49}},\ \bibinfo
  {pages} {259} (\bibinfo {year} {2006})}\BibitemShut {NoStop}%
\bibitem [{\citenamefont {Rogers}(2003)}]{R3}%
  \BibitemOpen
  \bibfield  {author} {\bibinfo {author} {\bibfnamefont {E.~M.}\ \bibnamefont
  {Rogers}},\ }\href@noop {} {\emph {\bibinfo {title} {Diffusion of
  Innovations}}},\ \bibinfo {edition} {5th}\ ed.\ (\bibinfo  {publisher} {Free
  Press},\ \bibinfo {address} {New York},\ \bibinfo {year} {2003})\BibitemShut
  {NoStop}%
\bibitem [{\citenamefont {Latora}\ and\ \citenamefont {Marchiori}(2004)}]{LM4}%
  \BibitemOpen
  \bibfield  {author} {\bibinfo {author} {\bibfnamefont {V.}~\bibnamefont
  {Latora}}\ and\ \bibinfo {author} {\bibfnamefont {M.}~\bibnamefont
  {Marchiori}},\ }\href@noop {} {\bibfield  {journal} {\bibinfo  {journal}
  {Chaos, Solitons, and Fractals},\ }\textbf {\bibinfo {volume} {20}},\
  \bibinfo {pages} {69} (\bibinfo {year} {2004})}\BibitemShut {NoStop}%
\bibitem [{\citenamefont {Skyrms}\ and\ \citenamefont {Pemantle}(2000)}]{SP0}%
  \BibitemOpen
  \bibfield  {author} {\bibinfo {author} {\bibfnamefont {B.}~\bibnamefont
  {Skyrms}}\ and\ \bibinfo {author} {\bibfnamefont {R.}~\bibnamefont
  {Pemantle}},\ }\href@noop {} {\bibfield  {journal} {\bibinfo  {journal}
  {Proc. Nat. Acad. Sci. USA},\ }\textbf {\bibinfo {volume} {97}},\ \bibinfo
  {pages} {9340} (\bibinfo {year} {2000})}\BibitemShut {NoStop}%
\bibitem [{\citenamefont {Doreian}\ and\ \citenamefont {Stokman}(1997)}]{DS97}%
  \BibitemOpen
  \bibinfo {editor} {\bibfnamefont {P.}~\bibnamefont {Doreian}}\ and\ \bibinfo
  {editor} {\bibfnamefont {F.~N.}\ \bibnamefont {Stokman}},\ eds.,\ \href@noop
  {} {\emph {\bibinfo {title} {Evolution of Social Networks}}}\ (\bibinfo
  {publisher} {Gordon and Breach},\ \bibinfo {address} {New York},\ \bibinfo
  {year} {1997})\BibitemShut {NoStop}%
\bibitem [{\citenamefont {Snijders}(2007)}]{S7}%
  \BibitemOpen
  \bibfield  {author} {\bibinfo {author} {\bibfnamefont {T.~A.~B.}\
  \bibnamefont {Snijders}},\ }in\ \href@noop {} {\emph {\bibinfo {booktitle}
  {Models and Methods in Social Network Analysis}}},\ \bibinfo {editor} {edited
  by\ \bibinfo {editor} {\bibfnamefont {P.~J.}\ \bibnamefont {Carrington}},
  \bibinfo {editor} {\bibfnamefont {J.}~\bibnamefont {Scott}}, \ and\ \bibinfo
  {editor} {\bibfnamefont {S.}~\bibnamefont {Wasserman}}}\ (\bibinfo
  {publisher} {Cambridge University Press},\ \bibinfo {address} {Cambridge,
  UK},\ \bibinfo {year} {2007})\BibitemShut {NoStop}%
\bibitem [{\citenamefont {Katz}\ and\ \citenamefont {Proctor}(1959)}]{KP59}%
  \BibitemOpen
  \bibfield  {author} {\bibinfo {author} {\bibfnamefont {L.}~\bibnamefont
  {Katz}}\ and\ \bibinfo {author} {\bibfnamefont {C.~H.}\ \bibnamefont
  {Proctor}},\ }\href@noop {} {\bibfield  {journal} {\bibinfo  {journal}
  {Psychometrika},\ }\textbf {\bibinfo {volume} {24}},\ \bibinfo {pages} {317}
  (\bibinfo {year} {1959})}\BibitemShut {NoStop}%
\bibitem [{\citenamefont {Wasserman}\ and\ \citenamefont
  {Iacobucci}(1988)}]{WI88}%
  \BibitemOpen
  \bibfield  {author} {\bibinfo {author} {\bibfnamefont {S.}~\bibnamefont
  {Wasserman}}\ and\ \bibinfo {author} {\bibfnamefont {D.}~\bibnamefont
  {Iacobucci}},\ }\href@noop {} {\bibfield  {journal} {\bibinfo  {journal}
  {Psychometrika},\ }\textbf {\bibinfo {volume} {53}},\ \bibinfo {pages} {261}
  (\bibinfo {year} {1988})}\BibitemShut {NoStop}%
\bibitem [{\citenamefont {Banks}\ and\ \citenamefont {Carley}(1996)}]{BC96}%
  \BibitemOpen
  \bibfield  {author} {\bibinfo {author} {\bibfnamefont {D.~L.}\ \bibnamefont
  {Banks}}\ and\ \bibinfo {author} {\bibfnamefont {K.~M.}\ \bibnamefont
  {Carley}},\ }\href@noop {} {\bibfield  {journal} {\bibinfo  {journal} {J Math
  Soc},\ }\textbf {\bibinfo {volume} {21}},\ \bibinfo {pages} {173} (\bibinfo
  {year} {1996})}\BibitemShut {NoStop}%
\bibitem [{\citenamefont {Robins}\ and\ \citenamefont {Pattison}(2001)}]{RP1}%
  \BibitemOpen
  \bibfield  {author} {\bibinfo {author} {\bibfnamefont {G.}~\bibnamefont
  {Robins}}\ and\ \bibinfo {author} {\bibfnamefont {P.}~\bibnamefont
  {Pattison}},\ }\href@noop {} {\bibfield  {journal} {\bibinfo  {journal} {J
  Math Soc},\ }\textbf {\bibinfo {volume} {25}},\ \bibinfo {pages} {5}
  (\bibinfo {year} {2001})}\BibitemShut {NoStop}%
\bibitem [{\citenamefont {Holland}\ and\ \citenamefont
  {Leinhardt}(1977)}]{HL77}%
  \BibitemOpen
  \bibfield  {author} {\bibinfo {author} {\bibfnamefont {P.~W.}\ \bibnamefont
  {Holland}}\ and\ \bibinfo {author} {\bibfnamefont {S.}~\bibnamefont
  {Leinhardt}},\ }\href@noop {} {\bibfield  {journal} {\bibinfo  {journal} {J
  Math Soc},\ }\textbf {\bibinfo {volume} {5}},\ \bibinfo {pages} {5} (\bibinfo
  {year} {1977})}\BibitemShut {NoStop}%
\bibitem [{\citenamefont {Wasserman}(1980)}]{W80}%
  \BibitemOpen
  \bibfield  {author} {\bibinfo {author} {\bibfnamefont {S.}~\bibnamefont
  {Wasserman}},\ }\href@noop {} {\bibfield  {journal} {\bibinfo  {journal} {J
  Amer Stat Assoc},\ }\textbf {\bibinfo {volume} {75}},\ \bibinfo {pages} {280}
  (\bibinfo {year} {1980})}\BibitemShut {NoStop}%
\bibitem [{\citenamefont {Snijders}(2001)}]{S1}%
  \BibitemOpen
  \bibfield  {author} {\bibinfo {author} {\bibfnamefont {T.~A.~B.}\
  \bibnamefont {Snijders}},\ }\href@noop {} {\bibfield  {journal} {\bibinfo
  {journal} {Sociological Methodology},\ }\textbf {\bibinfo {volume} {31}},\
  \bibinfo {pages} {361} (\bibinfo {year} {2001})}\BibitemShut {NoStop}%
\bibitem [{\citenamefont {Ahn}\ \emph {et~al.}(2007)\citenamefont {Ahn},
  \citenamefont {Han}, \citenamefont {Kwak}, \citenamefont {Moon},\ and\
  \citenamefont {Jeong}}]{AHKJ7}%
  \BibitemOpen
  \bibfield  {author} {\bibinfo {author} {\bibfnamefont {Y.~Y.}\ \bibnamefont
  {Ahn}}, \bibinfo {author} {\bibfnamefont {S.}~\bibnamefont {Han}}, \bibinfo
  {author} {\bibfnamefont {H.}~\bibnamefont {Kwak}}, \bibinfo {author}
  {\bibfnamefont {S.}~\bibnamefont {Moon}}, \ and\ \bibinfo {author}
  {\bibfnamefont {H.}~\bibnamefont {Jeong}},\ }in\ \href@noop {} {\emph
  {\bibinfo {booktitle} {WWW '07: Proceedings of the 16th international
  conference on World Wide Web}}}\ (\bibinfo {address} {New York},\ \bibinfo
  {year} {2007})\ pp.\ \bibinfo {pages} {835--844}\BibitemShut {NoStop}%
\bibitem [{\citenamefont {Holme}\ \emph {et~al.}(2004)\citenamefont {Holme},
  \citenamefont {Edling},\ and\ \citenamefont {Liljeros}}]{HEL4}%
  \BibitemOpen
  \bibfield  {author} {\bibinfo {author} {\bibfnamefont {P.}~\bibnamefont
  {Holme}}, \bibinfo {author} {\bibfnamefont {C.~R.}\ \bibnamefont {Edling}}, \
  and\ \bibinfo {author} {\bibfnamefont {F.}~\bibnamefont {Liljeros}},\
  }\href@noop {} {\bibfield  {journal} {\bibinfo  {journal} {Social Networks},\
  }\textbf {\bibinfo {volume} {26}},\ \bibinfo {pages} {155} (\bibinfo {year}
  {2004})}\BibitemShut {NoStop}%
\bibitem [{\citenamefont {Gross}\ and\ \citenamefont {Sayama}(2009)}]{GS9}%
  \BibitemOpen
  \bibinfo {editor} {\bibfnamefont {T.}~\bibnamefont {Gross}}\ and\ \bibinfo
  {editor} {\bibfnamefont {H.}~\bibnamefont {Sayama}},\ eds.,\ \href@noop {}
  {\emph {\bibinfo {title} {Adaptive Networks: Theory, Models and
  Applications}}}\ (\bibinfo  {publisher} {Springer},\ \bibinfo {address}
  {Heidelberg},\ \bibinfo {year} {2009})\BibitemShut {NoStop}%
\bibitem [{\citenamefont {Gross}\ and\ \citenamefont {Blasius}(2008)}]{GB8}%
  \BibitemOpen
  \bibfield  {author} {\bibinfo {author} {\bibfnamefont {T.}~\bibnamefont
  {Gross}}\ and\ \bibinfo {author} {\bibfnamefont {B.}~\bibnamefont
  {Blasius}},\ }\href@noop {} {\bibfield  {journal} {\bibinfo  {journal}
  {Journal of the Royal Society Interface},\ }\textbf {\bibinfo {volume} {5}},\
  \bibinfo {pages} {259} (\bibinfo {year} {2008})}\BibitemShut {NoStop}%
\bibitem [{\citenamefont {Bornholdt}\ and\ \citenamefont {Rohlf}(2000)}]{BR0}%
  \BibitemOpen
  \bibfield  {author} {\bibinfo {author} {\bibfnamefont {S.}~\bibnamefont
  {Bornholdt}}\ and\ \bibinfo {author} {\bibfnamefont {T.}~\bibnamefont
  {Rohlf}},\ }\href@noop {} {\bibfield  {journal} {\bibinfo  {journal} {Phys.
  Rev. Lett.},\ }\textbf {\bibinfo {volume} {84}},\ \bibinfo {pages} {6114}
  (\bibinfo {year} {2000})}\BibitemShut {NoStop}%
\bibitem [{\citenamefont {Holme}\ and\ \citenamefont {Ghoshal}(2006)}]{HG6}%
  \BibitemOpen
  \bibfield  {author} {\bibinfo {author} {\bibfnamefont {P.}~\bibnamefont
  {Holme}}\ and\ \bibinfo {author} {\bibfnamefont {G.}~\bibnamefont
  {Ghoshal}},\ }\href@noop {} {\bibfield  {journal} {\bibinfo  {journal} {Phys.
  Rev. Lett.},\ }\textbf {\bibinfo {volume} {96}},\ \bibinfo {pages} {098701}
  (\bibinfo {year} {2006})}\BibitemShut {NoStop}%
\bibitem [{\citenamefont {Ito}\ and\ \citenamefont {Kaneko}(2002)}]{IK2}%
  \BibitemOpen
  \bibfield  {author} {\bibinfo {author} {\bibfnamefont {J.}~\bibnamefont
  {Ito}}\ and\ \bibinfo {author} {\bibfnamefont {K.}~\bibnamefont {Kaneko}},\
  }\href@noop {} {\bibfield  {journal} {\bibinfo  {journal} {Phys. Rev.
  Lett.},\ }\textbf {\bibinfo {volume} {88}},\ \bibinfo {pages} {028701}
  (\bibinfo {year} {2002})}\BibitemShut {NoStop}%
\bibitem [{\citenamefont {Gil}\ and\ \citenamefont {Zanette}(2006)}]{GZ6}%
  \BibitemOpen
  \bibfield  {author} {\bibinfo {author} {\bibfnamefont {S.}~\bibnamefont
  {Gil}}\ and\ \bibinfo {author} {\bibfnamefont {D.~H.}\ \bibnamefont
  {Zanette}},\ }\href@noop {} {\bibfield  {journal} {\bibinfo  {journal} {Phys.
  Lett. A},\ }\textbf {\bibinfo {volume} {356}},\ \bibinfo {pages} {89}
  (\bibinfo {year} {2006})}\BibitemShut {NoStop}%
\bibitem [{\citenamefont {Gross}\ \emph {et~al.}(2006)\citenamefont {Gross},
  \citenamefont {D'Lima},\ and\ \citenamefont {Blasius}}]{GDB6}%
  \BibitemOpen
  \bibfield  {author} {\bibinfo {author} {\bibfnamefont {T.}~\bibnamefont
  {Gross}}, \bibinfo {author} {\bibfnamefont {C.~D.}\ \bibnamefont {D'Lima}}, \
  and\ \bibinfo {author} {\bibfnamefont {B.}~\bibnamefont {Blasius}},\
  }\href@noop {} {\bibfield  {journal} {\bibinfo  {journal} {Phys. Rev.
  Lett.},\ }\textbf {\bibinfo {volume} {96}},\ \bibinfo {pages} {208701}
  (\bibinfo {year} {2006})}\BibitemShut {NoStop}%
\bibitem [{\citenamefont {Shaw}\ and\ \citenamefont {Schwartz}(2008)}]{SS8}%
  \BibitemOpen
  \bibfield  {author} {\bibinfo {author} {\bibfnamefont {L.~B.}\ \bibnamefont
  {Shaw}}\ and\ \bibinfo {author} {\bibfnamefont {I.~B.}\ \bibnamefont
  {Schwartz}},\ }\href@noop {} {\bibfield  {journal} {\bibinfo  {journal}
  {Phys. Rev. E},\ }\textbf {\bibinfo {volume} {77}},\ \bibinfo {pages}
  {066101} (\bibinfo {year} {2008})}\BibitemShut {NoStop}%
\bibitem [{\citenamefont {Zimmerman}\ \emph {et~al.}(2004)\citenamefont
  {Zimmerman}, \citenamefont {Eguiluz},\ and\ \citenamefont {Miguel}}]{ZES4}%
  \BibitemOpen
  \bibfield  {author} {\bibinfo {author} {\bibfnamefont {M.~G.}\ \bibnamefont
  {Zimmerman}}, \bibinfo {author} {\bibfnamefont {V.~M.}\ \bibnamefont
  {Eguiluz}}, \ and\ \bibinfo {author} {\bibfnamefont {M.~S.}\ \bibnamefont
  {Miguel}},\ }\href@noop {} {\bibfield  {journal} {\bibinfo  {journal} {Phys.
  Rev. E},\ }\textbf {\bibinfo {volume} {69}},\ \bibinfo {pages} {065102}
  (\bibinfo {year} {2004})}\BibitemShut {NoStop}%
\bibitem [{\citenamefont {Pacheco}\ \emph {et~al.}(2006)\citenamefont
  {Pacheco}, \citenamefont {Traulsen},\ and\ \citenamefont {Nowak}}]{PTN6}%
  \BibitemOpen
  \bibfield  {author} {\bibinfo {author} {\bibfnamefont {J.~M.}\ \bibnamefont
  {Pacheco}}, \bibinfo {author} {\bibfnamefont {A.}~\bibnamefont {Traulsen}}, \
  and\ \bibinfo {author} {\bibfnamefont {M.~A.}\ \bibnamefont {Nowak}},\
  }\href@noop {} {\bibfield  {journal} {\bibinfo  {journal} {Phys. Rev.
  Lett.},\ }\textbf {\bibinfo {volume} {97}},\ \bibinfo {pages} {258103}
  (\bibinfo {year} {2006})}\BibitemShut {NoStop}%
\bibitem [{\citenamefont {Braha}\ and\ \citenamefont {Bar-Yam}(2006)}]{BB6}%
  \BibitemOpen
  \bibfield  {author} {\bibinfo {author} {\bibfnamefont {D.}~\bibnamefont
  {Braha}}\ and\ \bibinfo {author} {\bibfnamefont {Y.}~\bibnamefont
  {Bar-Yam}},\ }\href@noop {} {\bibfield  {journal} {\bibinfo  {journal}
  {Complexity},\ }\textbf {\bibinfo {volume} {12}},\ \bibinfo {pages} {59}
  (\bibinfo {year} {2006})}\BibitemShut {NoStop}%
\bibitem [{\citenamefont {Braha}\ and\ \citenamefont {Bar-Yam}(2009)}]{BB9}%
  \BibitemOpen
  \bibfield  {author} {\bibinfo {author} {\bibfnamefont {D.}~\bibnamefont
  {Braha}}\ and\ \bibinfo {author} {\bibfnamefont {Y.}~\bibnamefont
  {Bar-Yam}},\ }in\ \href@noop {} {\emph {\bibinfo {booktitle} {Adaptive
  Networks: Theory, Models and Applications}}},\ \bibinfo {editor} {edited by\
  \bibinfo {editor} {\bibfnamefont {T.}~\bibnamefont {Gross}}\ and\ \bibinfo
  {editor} {\bibfnamefont {H.}~\bibnamefont {Sayama}}}\ (\bibinfo  {publisher}
  {Springer},\ \bibinfo {year} {2009})\BibitemShut {NoStop}%
\bibitem [{\citenamefont {Barab{\'a}si}(2005)}]{B5}%
  \BibitemOpen
  \bibfield  {author} {\bibinfo {author} {\bibfnamefont {A.-L.}\ \bibnamefont
  {Barab{\'a}si}},\ }\href@noop {} {\bibfield  {journal} {\bibinfo  {journal}
  {Nature},\ }\textbf {\bibinfo {volume} {435}},\ \bibinfo {pages} {207}
  (\bibinfo {year} {2005})}\BibitemShut {NoStop}%
\bibitem [{\citenamefont {Wiggins}\ \emph
  {et~al.}(2008){\natexlab{a}}\citenamefont {Wiggins}, \citenamefont
  {Howison},\ and\ \citenamefont {Crowston}}]{WHC8}%
  \BibitemOpen
  \bibfield  {author} {\bibinfo {author} {\bibfnamefont {A.}~\bibnamefont
  {Wiggins}}, \bibinfo {author} {\bibfnamefont {J.}~\bibnamefont {Howison}}, \
  and\ \bibinfo {author} {\bibfnamefont {K.}~\bibnamefont {Crowston}},\
  }\href@noop {} {\bibfield  {journal} {\bibinfo  {journal} {IFIP International
  Federation for Information Processing},\ }\textbf {\bibinfo {volume} {275}},\
  \bibinfo {pages} {131} (\bibinfo {year} {2008}{\natexlab{a}})}\BibitemShut
  {NoStop}%
\bibitem [{\citenamefont {Wiggins}\ \emph
  {et~al.}(2008){\natexlab{b}}\citenamefont {Wiggins}, \citenamefont
  {McQuaid},\ and\ \citenamefont {Adamic}}]{WMA8}%
  \BibitemOpen
  \bibfield  {author} {\bibinfo {author} {\bibfnamefont {A.}~\bibnamefont
  {Wiggins}}, \bibinfo {author} {\bibfnamefont {M.}~\bibnamefont {McQuaid}}, \
  and\ \bibinfo {author} {\bibfnamefont {L.}~\bibnamefont {Adamic}},\ }in\
  \href@noop {} {\emph {\bibinfo {booktitle} {iConference 2008: iFutures:
  Systems, Selves, Society}}}\ (\bibinfo {address} {Los Angeles},\ \bibinfo
  {year} {2008})\BibitemShut {NoStop}%
\bibitem [{\citenamefont {Adamic}\ \emph {et~al.}(2009)\citenamefont {Adamic},
  \citenamefont {Brunetti}, \citenamefont {Harris},\ and\ \citenamefont
  {Kirilenko}}]{ABH9}%
  \BibitemOpen
  \bibfield  {author} {\bibinfo {author} {\bibfnamefont {L.}~\bibnamefont
  {Adamic}}, \bibinfo {author} {\bibfnamefont {C.}~\bibnamefont {Brunetti}},
  \bibinfo {author} {\bibfnamefont {J.~H.}\ \bibnamefont {Harris}}, \ and\
  \bibinfo {author} {\bibfnamefont {A.~A.}\ \bibnamefont {Kirilenko}},\
  }\href@noop {} {\enquote {\bibinfo {title} {On the informational properties
  of trading networks},}\ }\bibinfo {howpublished} {Working Paper. School of
  Information \& Center for the Study of Complex Systems, University of
  Michigan} (\bibinfo {year} {2009})\BibitemShut {NoStop}%
\bibitem [{\citenamefont {Han}\ \emph {et~al.}(2004)\citenamefont {Han},
  \citenamefont {Bertin}, \citenamefont {Hao}, \citenamefont {Goldberg},
  \citenamefont {Berriz}, \citenamefont {Zhang}, \citenamefont {Dupuy},
  \citenamefont {Walhout}, \citenamefont {Cusick}, \citenamefont {Roth},\ and\
  \citenamefont {Vidal}}]{HBH4}%
  \BibitemOpen
  \bibfield  {author} {\bibinfo {author} {\bibfnamefont {J.~D.~J.}\
  \bibnamefont {Han}}, \bibinfo {author} {\bibfnamefont {N.}~\bibnamefont
  {Bertin}}, \bibinfo {author} {\bibfnamefont {T.}~\bibnamefont {Hao}},
  \bibinfo {author} {\bibfnamefont {D.~S.}\ \bibnamefont {Goldberg}}, \bibinfo
  {author} {\bibfnamefont {G.~F.}\ \bibnamefont {Berriz}}, \bibinfo {author}
  {\bibfnamefont {L.~V.}\ \bibnamefont {Zhang}}, \bibinfo {author}
  {\bibfnamefont {D.}~\bibnamefont {Dupuy}}, \bibinfo {author} {\bibfnamefont
  {A.~J.~M.}\ \bibnamefont {Walhout}}, \bibinfo {author} {\bibfnamefont
  {M.~E.}\ \bibnamefont {Cusick}}, \bibinfo {author} {\bibfnamefont {F.~P.}\
  \bibnamefont {Roth}}, \ and\ \bibinfo {author} {\bibfnamefont
  {M.}~\bibnamefont {Vidal}},\ }\href@noop {} {\bibfield  {journal} {\bibinfo
  {journal} {Nature},\ }\textbf {\bibinfo {volume} {430}},\ \bibinfo {pages}
  {88} (\bibinfo {year} {2004})}\BibitemShut {NoStop}%
\bibitem [{\citenamefont {Luscombe}\ \emph {et~al.}(2004)\citenamefont
  {Luscombe}, \citenamefont {Babu}, \citenamefont {Yu}, \citenamefont {Snyder},
  \citenamefont {Teichmann},\ and\ \citenamefont {Gerstein}}]{LBY4}%
  \BibitemOpen
  \bibfield  {author} {\bibinfo {author} {\bibfnamefont {N.~M.}\ \bibnamefont
  {Luscombe}}, \bibinfo {author} {\bibfnamefont {M.~M.}\ \bibnamefont {Babu}},
  \bibinfo {author} {\bibfnamefont {H.}~\bibnamefont {Yu}}, \bibinfo {author}
  {\bibfnamefont {M.}~\bibnamefont {Snyder}}, \bibinfo {author} {\bibfnamefont
  {S.~A.}\ \bibnamefont {Teichmann}}, \ and\ \bibinfo {author} {\bibfnamefont
  {M.}~\bibnamefont {Gerstein}},\ }\href@noop {} {\bibfield  {journal}
  {\bibinfo  {journal} {Nature},\ }\textbf {\bibinfo {volume} {431}},\ \bibinfo
  {pages} {308} (\bibinfo {year} {2004})}\BibitemShut {NoStop}%
\bibitem [{\citenamefont {Doyle}\ and\ \citenamefont {Snell}(1984)}]{DS84}%
  \BibitemOpen
  \bibfield  {author} {\bibinfo {author} {\bibfnamefont {P.~G.}\ \bibnamefont
  {Doyle}}\ and\ \bibinfo {author} {\bibfnamefont {J.~L.}\ \bibnamefont
  {Snell}},\ }\href@noop {} {\emph {\bibinfo {title} {Random Walks and Electric
  Networks}}}\ (\bibinfo  {publisher} {The Mathematical Association of
  America},\ \bibinfo {year} {1984})\BibitemShut {NoStop}%
\bibitem [{\citenamefont {Pemantle}(2007)}]{P7}%
  \BibitemOpen
  \bibfield  {author} {\bibinfo {author} {\bibfnamefont {R.}~\bibnamefont
  {Pemantle}},\ }\href@noop {} {\bibfield  {journal} {\bibinfo  {journal}
  {Probability Surveys},\ }\textbf {\bibinfo {volume} {4}},\ \bibinfo {pages}
  {1} (\bibinfo {year} {2007})}\BibitemShut {NoStop}%
\bibitem [{\citenamefont {Jespersen}\ and\ \citenamefont {Blumen}(2000)}]{JB0}%
  \BibitemOpen
  \bibfield  {author} {\bibinfo {author} {\bibfnamefont {S.}~\bibnamefont
  {Jespersen}}\ and\ \bibinfo {author} {\bibfnamefont {A.}~\bibnamefont
  {Blumen}},\ }\href@noop {} {\bibfield  {journal} {\bibinfo  {journal} {Phys.
  Rev. E},\ }\textbf {\bibinfo {volume} {62}},\ \bibinfo {pages} {6270}
  (\bibinfo {year} {2000})}\BibitemShut {NoStop}%
\bibitem [{\citenamefont {Jespersen}\ \emph {et~al.}(2000)\citenamefont
  {Jespersen}, \citenamefont {Sokolov},\ and\ \citenamefont {Blumen}}]{JSB0}%
  \BibitemOpen
  \bibfield  {author} {\bibinfo {author} {\bibfnamefont {S.}~\bibnamefont
  {Jespersen}}, \bibinfo {author} {\bibfnamefont {I.~M.}\ \bibnamefont
  {Sokolov}}, \ and\ \bibinfo {author} {\bibfnamefont {A.}~\bibnamefont
  {Blumen}},\ }\href@noop {} {\bibfield  {journal} {\bibinfo  {journal} {Phys.
  Rev. E},\ }\textbf {\bibinfo {volume} {62}},\ \bibinfo {pages} {4405}
  (\bibinfo {year} {2000})}\BibitemShut {NoStop}%
\bibitem [{\citenamefont {Lahtinen}\ \emph {et~al.}(2001)\citenamefont
  {Lahtinen}, \citenamefont {Kert{\'e}sz},\ and\ \citenamefont {Kaski}}]{LKK1}%
  \BibitemOpen
  \bibfield  {author} {\bibinfo {author} {\bibfnamefont {J.}~\bibnamefont
  {Lahtinen}}, \bibinfo {author} {\bibfnamefont {J.}~\bibnamefont
  {Kert{\'e}sz}}, \ and\ \bibinfo {author} {\bibfnamefont {K.}~\bibnamefont
  {Kaski}},\ }\href@noop {} {\bibfield  {journal} {\bibinfo  {journal} {Phys.
  Rev. E},\ }\textbf {\bibinfo {volume} {64}},\ \bibinfo {pages} {057105}
  (\bibinfo {year} {2001})}\BibitemShut {NoStop}%
\bibitem [{\citenamefont {Manrubia}\ \emph {et~al.}(2001)\citenamefont
  {Manrubia}, \citenamefont {Delgado},\ and\ \citenamefont {Luque}}]{MDL1}%
  \BibitemOpen
  \bibfield  {author} {\bibinfo {author} {\bibfnamefont {S.~C.}\ \bibnamefont
  {Manrubia}}, \bibinfo {author} {\bibfnamefont {J.}~\bibnamefont {Delgado}}, \
  and\ \bibinfo {author} {\bibfnamefont {B.}~\bibnamefont {Luque}},\
  }\href@noop {} {\bibfield  {journal} {\bibinfo  {journal} {Europhys. Lett.},\
  }\textbf {\bibinfo {volume} {53}},\ \bibinfo {pages} {693} (\bibinfo {year}
  {2001})}\BibitemShut {NoStop}%
\bibitem [{\citenamefont {Pandit}\ and\ \citenamefont {Amritkar}(2001)}]{PA1}%
  \BibitemOpen
  \bibfield  {author} {\bibinfo {author} {\bibfnamefont {S.~A.}\ \bibnamefont
  {Pandit}}\ and\ \bibinfo {author} {\bibfnamefont {R.~E.}\ \bibnamefont
  {Amritkar}},\ }\href@noop {} {\bibfield  {journal} {\bibinfo  {journal}
  {Phys. Rev. E},\ }\textbf {\bibinfo {volume} {63}},\ \bibinfo {pages}
  {041104} (\bibinfo {year} {2001})}\BibitemShut {NoStop}%
\bibitem [{\citenamefont {Jasch}\ and\ \citenamefont {Blumen}(2001)}]{JB1}%
  \BibitemOpen
  \bibfield  {author} {\bibinfo {author} {\bibfnamefont {F.}~\bibnamefont
  {Jasch}}\ and\ \bibinfo {author} {\bibfnamefont {A.}~\bibnamefont {Blumen}},\
  }\href@noop {} {\bibfield  {journal} {\bibinfo  {journal} {Phys. Rev. E},\
  }\textbf {\bibinfo {volume} {64}},\ \bibinfo {pages} {066104} (\bibinfo
  {year} {2001})}\BibitemShut {NoStop}%
\bibitem [{\citenamefont {Almaas}\ \emph {et~al.}(2002)\citenamefont {Almaas},
  \citenamefont {Kulkarni},\ and\ \citenamefont {Stroud}}]{AKS2}%
  \BibitemOpen
  \bibfield  {author} {\bibinfo {author} {\bibfnamefont {E.}~\bibnamefont
  {Almaas}}, \bibinfo {author} {\bibfnamefont {R.~V.}\ \bibnamefont
  {Kulkarni}}, \ and\ \bibinfo {author} {\bibfnamefont {D.}~\bibnamefont
  {Stroud}},\ }\href@noop {} {\bibfield  {journal} {\bibinfo  {journal} {Phys.
  Rev. Lett.},\ }\textbf {\bibinfo {volume} {88}},\ \bibinfo {pages} {098101}
  (\bibinfo {year} {2002})}\BibitemShut {NoStop}%
\bibitem [{\citenamefont {Luque}\ and\ \citenamefont {Miramontes}(2003)}]{LM3}%
  \BibitemOpen
  \bibfield  {author} {\bibinfo {author} {\bibfnamefont {B.}~\bibnamefont
  {Luque}}\ and\ \bibinfo {author} {\bibfnamefont {O.}~\bibnamefont
  {Miramontes}},\ }\href@noop {} {\bibfield  {journal} {\bibinfo  {journal}
  {Europhys. Lett.},\ }\textbf {\bibinfo {volume} {63}},\ \bibinfo {pages} {8}
  (\bibinfo {year} {2003})}\BibitemShut {NoStop}%
\bibitem [{\citenamefont {Adamic}\ \emph {et~al.}(2001)\citenamefont {Adamic},
  \citenamefont {Lukose}, \citenamefont {Puniyani},\ and\ \citenamefont
  {Huberman}}]{ALP1}%
  \BibitemOpen
  \bibfield  {author} {\bibinfo {author} {\bibfnamefont {L.}~\bibnamefont
  {Adamic}}, \bibinfo {author} {\bibfnamefont {R.~M.}\ \bibnamefont {Lukose}},
  \bibinfo {author} {\bibfnamefont {A.~R.}\ \bibnamefont {Puniyani}}, \ and\
  \bibinfo {author} {\bibfnamefont {B.~A.}\ \bibnamefont {Huberman}},\
  }\href@noop {} {\bibfield  {journal} {\bibinfo  {journal} {Phys. Rev. E},\
  }\textbf {\bibinfo {volume} {64}},\ \bibinfo {pages} {046135} (\bibinfo
  {year} {2001})}\BibitemShut {NoStop}%
\bibitem [{\citenamefont {Newman}\ and\ \citenamefont {Girvan}(2004)}]{NG4}%
  \BibitemOpen
  \bibfield  {author} {\bibinfo {author} {\bibfnamefont {M.~E.~J.}\
  \bibnamefont {Newman}}\ and\ \bibinfo {author} {\bibfnamefont
  {M.}~\bibnamefont {Girvan}},\ }\href@noop {} {\bibfield  {journal} {\bibinfo
  {journal} {Phys. Rev. E},\ }\textbf {\bibinfo {volume} {69}},\ \bibinfo
  {pages} {026113} (\bibinfo {year} {2004})}\BibitemShut {NoStop}%
\bibitem [{\citenamefont {Bilke}\ and\ \citenamefont {Peterson}(2001)}]{BP1}%
  \BibitemOpen
  \bibfield  {author} {\bibinfo {author} {\bibfnamefont {S.}~\bibnamefont
  {Bilke}}\ and\ \bibinfo {author} {\bibfnamefont {C.}~\bibnamefont
  {Peterson}},\ }\href@noop {} {\bibfield  {journal} {\bibinfo  {journal}
  {Phys. Rev. E},\ }\textbf {\bibinfo {volume} {64}},\ \bibinfo {pages}
  {036106} (\bibinfo {year} {2001})}\BibitemShut {NoStop}%
\bibitem [{\citenamefont {da~Fontoura~Costa}\ and\ \citenamefont
  {Travieso}(2007)}]{CT7}%
  \BibitemOpen
  \bibfield  {author} {\bibinfo {author} {\bibfnamefont {L.}~\bibnamefont
  {da~Fontoura~Costa}}\ and\ \bibinfo {author} {\bibfnamefont {G.}~\bibnamefont
  {Travieso}},\ }\href@noop {} {\bibfield  {journal} {\bibinfo  {journal}
  {Phys. Rev. E},\ }\textbf {\bibinfo {volume} {75}},\ \bibinfo {pages}
  {016102} (\bibinfo {year} {2007})}\BibitemShut {NoStop}%
\bibitem [{\citenamefont {da~Fontoura~Costa}\ \emph
  {et~al.}(2007){\natexlab{b}}\citenamefont {da~Fontoura~Costa}, \citenamefont
  {Sporns}, \citenamefont {Antiqueira}, \citenamefont {Nunes},\ and\
  \citenamefont {Jr}}]{CSA7}%
  \BibitemOpen
  \bibfield  {author} {\bibinfo {author} {\bibfnamefont {L.}~\bibnamefont
  {da~Fontoura~Costa}}, \bibinfo {author} {\bibfnamefont {O.}~\bibnamefont
  {Sporns}}, \bibinfo {author} {\bibfnamefont {L.}~\bibnamefont {Antiqueira}},
  \bibinfo {author} {\bibfnamefont {M.~G.~V.}\ \bibnamefont {Nunes}}, \ and\
  \bibinfo {author} {\bibfnamefont {O.~N.~O.}\ \bibnamefont {Jr}},\ }\href@noop
  {} {\bibfield  {journal} {\bibinfo  {journal} {App. Phys. Lett.},\ }\textbf
  {\bibinfo {volume} {91}},\ \bibinfo {pages} {054107} (\bibinfo {year}
  {2007}{\natexlab{b}})}\BibitemShut {NoStop}%
\bibitem [{Note1()}]{Note1}%
  \BibitemOpen
  \bibinfo {note} {The underlay is a standard Barabasi-Albert network: the
  exponent is not the familiar $\alpha =3$ because we are dealing with the
  \protect \emph {cumulative} degree distribution, which decreases the exponent
  by one.}\BibitemShut {Stop}%
\end{thebibliography}
\end{document}